\documentclass[aps,superscriptaddress,twocolumn,twoside,floatfix,pra,nofootinbib,a4paper]{revtex4-2}
\usepackage{times}
\usepackage{amsmath,amssymb,amsthm,mathtools,thmtools,nameref}
\usepackage{comment}
\usepackage{enumerate}
\usepackage{float}
\usepackage[dvipsnames]{xcolor}
\usepackage{xcolor}
\usepackage[colorlinks=true,linkcolor=blue,citecolor=magenta,urlcolor=blue]{hyperref}
\usepackage[framemethod=default]{mdframed}
\usepackage{csvsimple}
\usepackage[mode=tex]{standalone}
\usepackage[capitalize]{cleveref}
\usepackage{multirow}
\usepackage{booktabs}
\usepackage{quantikz}
\usepackage{colortbl}

\newcommand{\tr}{\text{Tr}}

\begin{document}
\title{Measurement-Based Long-Range Entangling Gates in Constant Depth}
\author{Elisa~B\"aumer}
\email{eba@zurich.ibm.com}
\affiliation{IBM Quantum, IBM Research -- Zurich, 8803 R\"uschlikon, Switzerland}
\author{Stefan~Woerner}
\affiliation{IBM Quantum, IBM Research -- Zurich, 8803 R\"uschlikon, Switzerland}
\date{\today}%

\begin{abstract}
The depth of quantum circuits is a critical factor when running them on state-of-the-art quantum devices due to their limited coherence times. Reducing circuit depth decreases noise in near-term quantum computations and reduces overall computation time, thus, also benefiting fault-tolerant quantum computations. Here, we show how to reduce the depth of quantum sub-routines that typically scale linearly with the number of qubits, such as quantum fan-out and long-range CNOT gates, to a constant depth using mid-circuit measurements and feed-forward operations, while only requiring a 1D line topology. We compare our protocols with existing ones to highlight their advantages. Additionally, we verify the feasibility by implementing the measurement-based quantum fan-out gate and long-range CNOT gate on real quantum hardware, demonstrating significant improvements over their unitary implementations.
\end{abstract}

\maketitle

\section{Introduction}
As quantum computers are scaling up and we are entering the era of quantum utility \cite{kim2023utility}, the depth of quantum circuits that entangle all qubits naturally increases. Due to limited coherence times, this leads to more noise in near-term quantum computations. However, even for fault-tolerant quantum computations, reducing the depth is desirable as it simply allows quantum algorithms to run faster.

Recent advances in quantum technologies and the capability to implement dynamic circuits, i.e., quantum circuits enhanced with mid-circuit measurements and real-time feed-forward operations conditioned on classical calculations based on these mid-circuit measurement outcomes, also known as local alternating quantum classical computations (LAQCC) circuits, led to increasing interest in such circuits and measurement-based quantum algorithms.
A number of projects focused on mid-circuit measurements and feed-forward operations to reduce the depth of state preparation~\cite{Malz2024,sahay2024finitedepthpreparationtensornetwork,stephen2024preparingmatrixproductstates,zhang2024characterizingmpspepspreparable}, as well as the implementations of such protocols~\cite{iqbal2023,fossfeig2023,baeumer2023,chen2023nishimori}. 

Here, we want to use dynamic circuits to reduce the depth of quantum sub-routines that are not necessarily at the beginning of a circuit. Twenty years ago, Josza already derived a teleportation-based protocol to implement Clifford circuits in constant depth~\cite{jozsa2005introduction}. In the worst case, this protocol requires quadratically many ancilla qubits and a 2D connectivity, and, thus, is in general impractical.
Recently, \cite{buhrman2023state} reconsidered this approach and showed that for the quantum fan-out gate $2n$ ancilla qubits and a 1D line connectivity are sufficient. 
In~\cite{piroli2024approximating} another method to implement the quantum fan-out gate in constant-depth has been proposed that requires only $n$ ancilla qubits. However, this approach requires a ``ladder'' connectivity. For devices that do not offer this connectivity, such as the IBM Quantum devices with their heavy-hex topology~\cite{iqp}, this method becomes significantly more difficult to implement. 

Motivated by the ability of dynamic circuits to significantly reduce the circuit depth for specific quantum sub-routines, we explored further schemes to implement long-range entangling operations in constant depth. 
Our main results are introduced in Section~\ref{sec:ourimplementation}, where we introduce improved constant depth implementations of the following gates acting on $(n+1)$-qubit using only $n$ ancilla qubits and requiring only a 1D line topology: We show how to implement CNOT ladders on $(n+1)$ qubits, fan-out gates with one control and $n$ target qubits, long-range CNOT gates skipping $n-1$ (non-ancilla) qubits, as well as $(n+1)$-qubit qubit $R_{ZZ...Z}$ rotations and fan-out gates with $n$ arbitrary controlled-single-qubit unitaries.
Further, the new protocols are summarized in Tab.~\ref{tab:comparison}.
To demonstrate the improvement of our techniques to previous ones, we compare our protocol and the required resources for implementing the quantum fan-out gate with existing protocols in Section~\ref{sec:comparison}. We then implement the fan-out gate, as well as the long-range CNOT gate on quantum devices, substantiating the benefit of using measurement and feed-forward operations in Section~\ref{sec:experiments}. In Section~\ref{sec:applications} we describe how these sub-routines can be used for different applications. Finally, we conclude in Section~\ref{sec:conclusion}.

\section{Constant depth gate constructions}
\label{sec:ourimplementation}

In the following we present different constructions for quantum sub-routines in constant depth using mid-circuit measurements and feed-forward operations. Each of the following schemes can be implemented on $n+1$ system qubits alternating with $n$ additional ancilla qubits on a 1D line, which makes it particularly practical for implementations on state-of-the-art quantum devices.

\subsection{CNOT-ladders}
\label{sec:ladders}
\begin{figure}[htb]
\includegraphics[width=1.0\columnwidth]{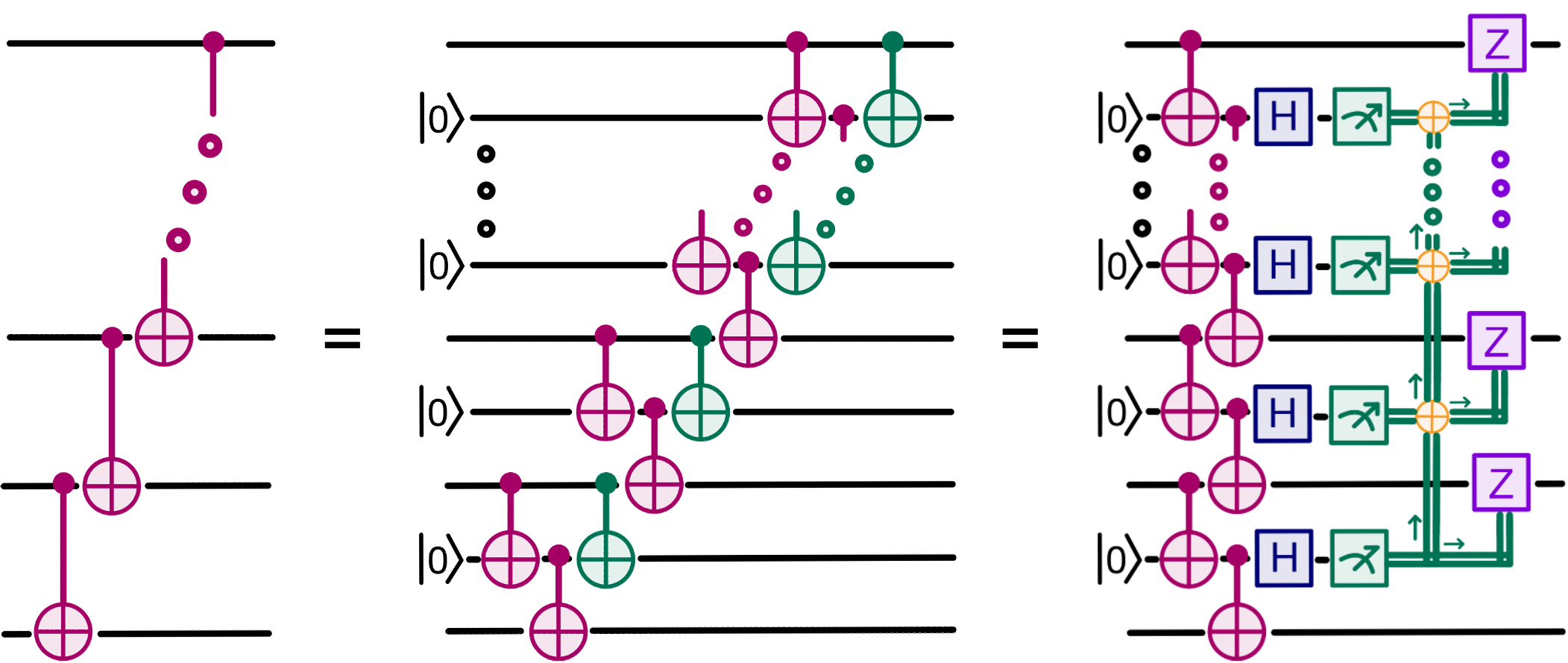}
\caption{CNOT ladders and their implementation using dynamic circuits (for the derivation see Appendix~\ref{app:derivation_ladders}).
} 
\label{fig:ladders}
\end{figure}

An efficient way to implement CNOT-ladders using dynamic circuits is illustrated in Fig.~\ref{fig:ladders}. Implementing a CNOT ladder consisting of $n$ CNOT gates in constant depth requires $n$ additional ancilla qubits and $n$ (parallel) mid-circuit measurements, followed by one round of feed-forward operations. Analogously, CNOT-ladders of opposite orientation can be implemented by exchanging the Z-basis and X-basis (i.e. applying Hadamard gates on all qubits before and after). For a more detailed derivation of the equivalence between unitary and dynamic circuits using commutation relations see Appendix~\ref{app:derivation_ladders}.

\subsection{Fan-out gate}
\label{sec:fanout}
\begin{figure}[htb]
\includegraphics[width=1.0\columnwidth]{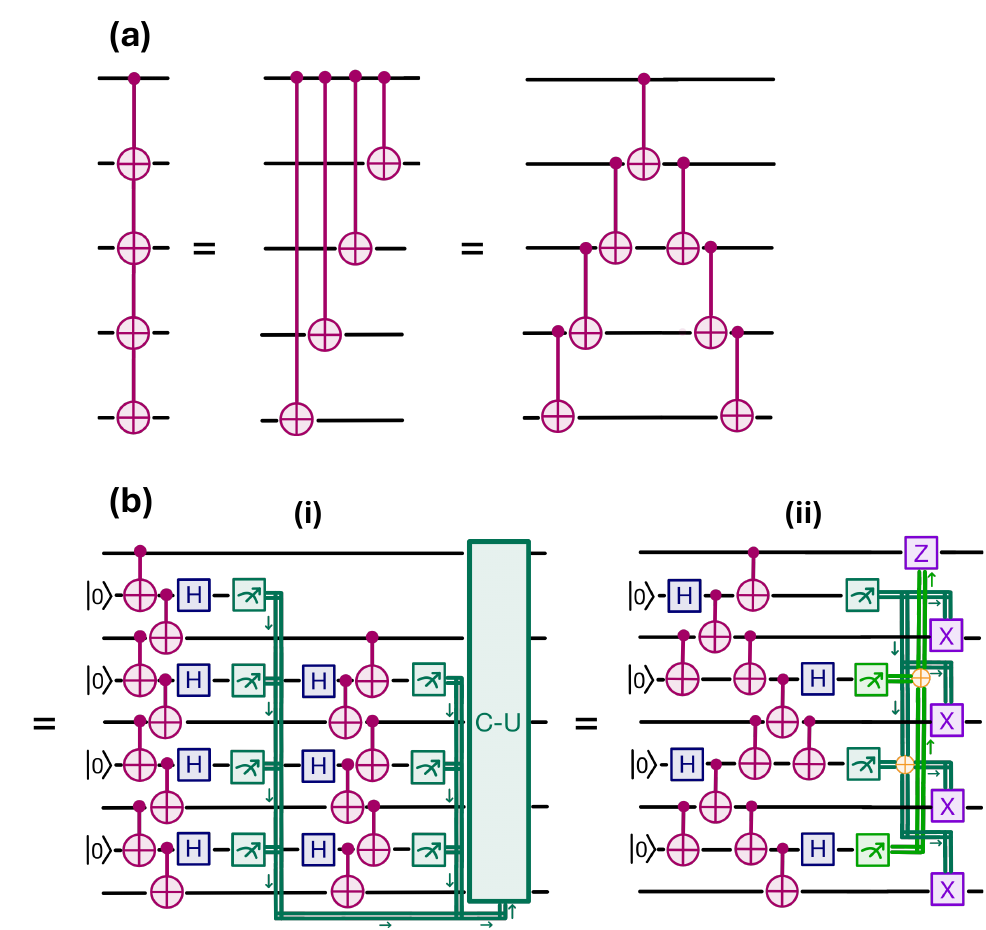}
\caption{(a) The quantum fan-out gate, here acting on $n=4$ target qubits. (b) Measurement-based implementation of the fan-out gate (i) by fusion of two CNOT-ladders and (ii) with reduction to only one round of mid-circuit measurements. 
} 
\label{fig:fanout_dyn}
\end{figure}
While the fan-in in classical computing corresponds to the number of inputs a logical gate can handle, the fan-out determines the number of new inputs a given logic output drives. We call a quantum fan-out gate on $n+1$ qubits a sequence of $n$ CNOT gates sharing one control qubit, as the information of the control qubit essentially gets copied into the other $n$ qubits, assuming they are in the empty state $\ket{0}$ before. It has been known for a long time that such quantum fan-out gates could be very powerful~\cite{hoyer2005}, by allowing for the implementation of many arithmetic operations, sub-routines and algorithms in constant depth. 

A quantum fan-out gate can be implemented using two CNOT ladders, see Fig.~\ref{fig:fanout_dyn}(a).
Thus, we can construct a fan-out gate in constant depth with $n$ ancilla qubits and $2n-1$ mid-circuit measurements. Instead of two rounds of classically conditioned feed-forward operations, we can use commutation relations to push the first ones through the circuit and implement it with only one round of feed-forward operations, as illustrated in Fig.~\ref{fig:fanout_dyn}(b)~(i). 
However, as mid-circuit measurements are comparatively costly because of the time they take and the amount of noise they introduce, we present a second implementation with dynamic circuits in Fig.~\ref{fig:fanout_dyn}(b)~(ii), that only requires one round of mid-circuit measurements and feed-forward operations and less CNOT gates in total. Its construction relies only on the circuit equivalences described in Appendix~\ref{app:derivations} and a detailed derivation can be found in Appendix~\ref{app:derivation_fanout}.

\subsection{Long-Range CNOT gate}
\label{sec:cnot}
\begin{figure}[htb]
\includegraphics[width=0.8\columnwidth]{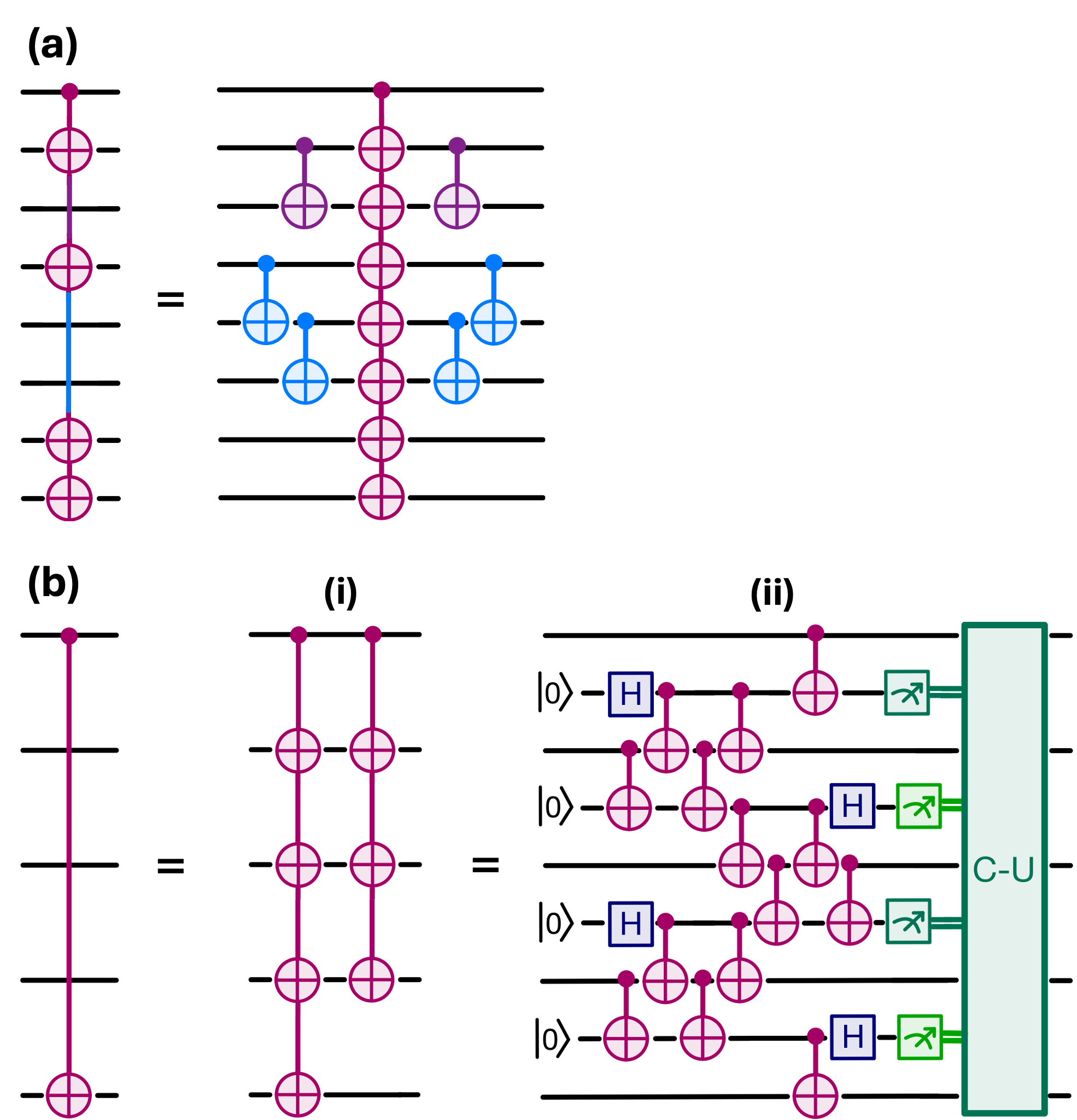}
\caption{(a) Skipping some qubits in the fan-out gate. (b) Implementation of the long-range CNOT gate in constant depth by (i) applying two fan-out gates of consecutive size and (ii) using essentially gate teleportation via ancilla qubits with reduction to only one round of mid-circuit measurements.
} 
\label{fig:fanout_skip}
\end{figure}
The fan-out gate does not require to act on all qubits, but its effect on individual qubits can essentially be ``skipped" by applying CNOT gates from the left and right, as illustrated in Fig.~\ref{fig:fanout_skip}(a). For skipping multiple qubits in a row, this yields CNOT ladders, which can be implemented in constant depth as well. In the most extreme case of unentangling all qubits along the way, this allows the implementation of a long-range CNOT gate in constant depth with three rounds of mid-circuit measurements. 

The long-range CNOT gate can also be implemented by applying two fan-out gates of consecutive size as shown in Fig.~\ref{fig:fanout_skip}(b)~(i), reducing the number of mid-circuit measurement rounds to two. However, we can reduce it even further to a single round of mid-circuit measurements and less CNOT gates, as shown in Fig.~\ref{fig:fanout_skip}(b)~(ii), using essentially gate teleportation via ancilla qubits and ``jumping" over the system qubits, as shown in Appendix~\ref{app:derivation_cnot}. 

While this would directly imply the implementation of a SWAP gate in constant depth with three rounds of mid-circuit measurements, we can implement it with just two rounds of mid-circuit measurements by essentially teleporting one qubit to an ancilla next to the other one as described in Appendix~\ref{app:derivation_teleportation}, performing a local SWAP gate, and then teleporting the swapped qubit back.

\subsection{Multi-qubit (parametrized) rotation gates}
\label{sec:arbitraryU}
\begin{figure}[htb]
\includegraphics[width=1.0\columnwidth]{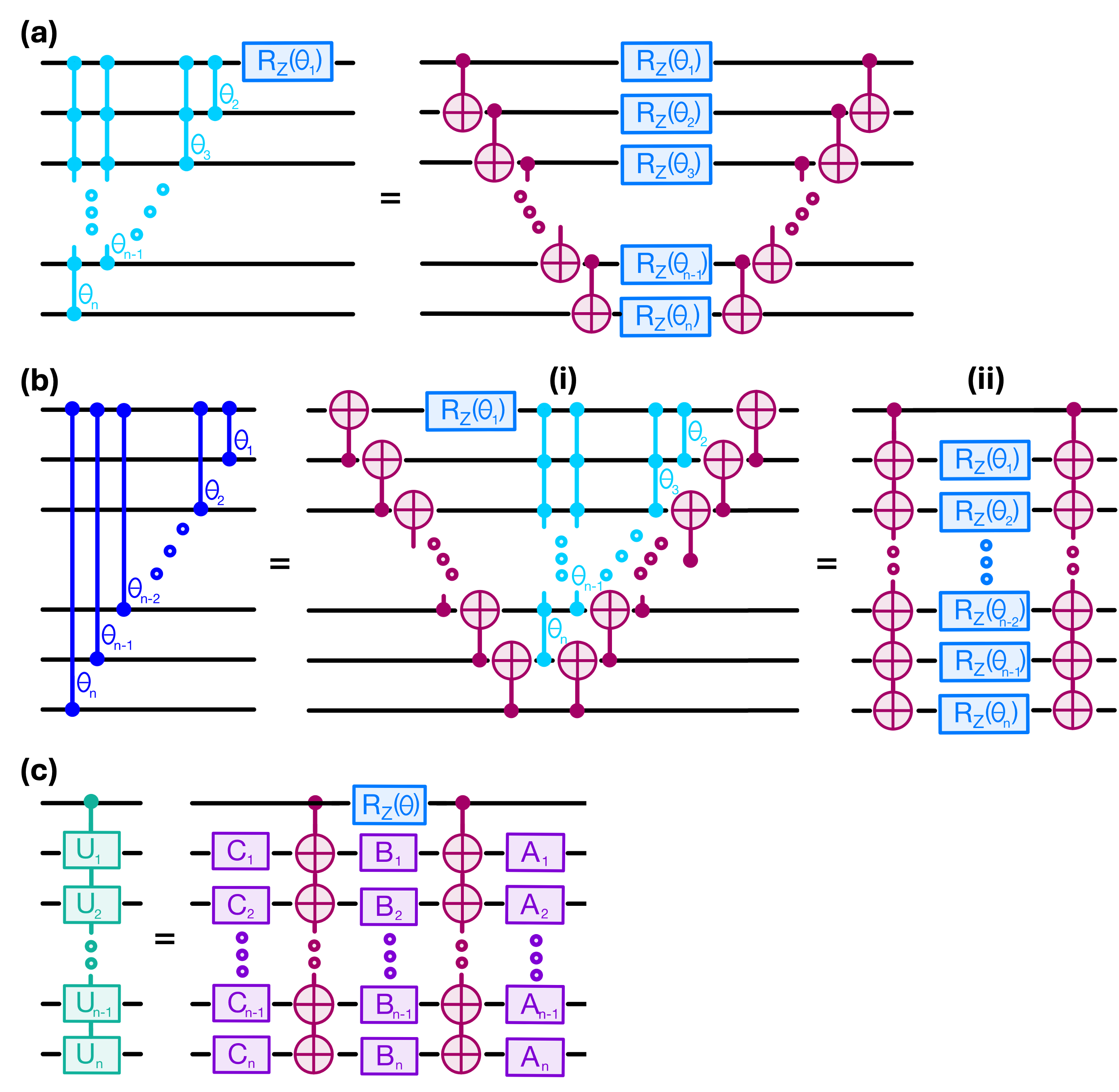}
\caption{(a) Construction of parallel multi-qubit $R_{ZZ...Z}$ gates from CNOT ladders. (b) $R_{ZZ}$ gate construction. (c) Fan-out with parallel single-qubit unitaries
} 
\label{fig:rzz}
\end{figure}

When sandwiching arbitrary phase rotations $R_z(\theta_k)$ on qubits $q_{k-1}$, $k=1..n$, between two CNOT ladders, the resulting operation is the product of $n$ Pauli rotations acting as joint Z-rotations on the first $k$ qubits with angles $\theta_k$, i.e.
$\prod_{k=1}^n R_{Z^{\otimes k}}(\theta_k)_{q_0..q_{k-1}}$, as shown in Fig.~\ref{fig:rzz}(a).
Sandwiching the resulting operation between another two CNOT ladders of opposite direction, as shown in Fig.~\ref{fig:rzz}(b)(i), the total operation is the product of $n$ $R_{ZZ}(\theta_k)$ gates acting on the same control qubit $q_0$ and the different qubits $q_k$, $k=1..n$, i.e.,
$\prod_{k=1}^n R_{ZZ}(\theta_k)_{q_0,q_k}$.
Another way to implement the same operation is by sandwiching single-qubit rotations $R_Z(\theta_k)$ on the different qubits $q_k$, $k=1..n$, as shown in Fig.~\ref{fig:rzz}(b)(ii). Last, using the fact that every controlled single-qubit rotation $U$ can be decomposed into two CNOT gates, one Z-rotation $R_Z(\theta)$ and three arbitrary single-qubit rotations $A, B, C$, such that $ABC=\mathbb{I}$ and $U=e^{i\theta}AXBXC$ (where $X$ is the bit-flip Pauli operator)~\cite{NielsenChuang}, we can decompose any fan-out gate with arbitrary parallel single-qubit unitaries controlled by the same qubit as shown in Fig.~\ref{fig:rzz}(c).

\section{Comparison with existing protocols}
\label{sec:comparison}

\begin{table*}[htb]
    \centering
    \begin{ruledtabular}
    \begin{tabular}{ll|cccccc}
        \multirow{2}{*}{Sub-routine} & \multirow{2}{*}{Implementation}& \multirow{2}{*}{Connectivity} & \multirow{2}{*}{\# qubits}  & rounds of mid-circuit & \# mid-circuit & depth of & \multirow{2}{*}{\# CNOT gates} \\ 
        &  &  & & measurements & measurements & CNOT layers &  \\ \toprule
        \rowcolor{violet!4}[3.5\tabcolsep] & & star &  & &  & $n$ & $n$ \\
        \rowcolor{violet!4}[3.5\tabcolsep] \multirow{-2}{*} {Fanout} & \multirow{-2}{*} {Unitary} & 1D line &  \multirow{-2}{*}{$n+1$} & \multirow{-2}{*}{$0$}  & \multirow{-2}{*}{$0$} & $2n-1$ & $2n-1$ \\ \hline
        \rowcolor{violet!12}[3.5\tabcolsep] Fanout & Buhrman et al.~\cite{buhrman2023state} & 1D line & $3n+1$ & $2$ & $4n$ & $6$ & $6n-1$ \\ \hline
        \rowcolor{violet!12}[3.5\tabcolsep]  & & ladder & & & & $4$ & $\frac{5}{2}n-2$\\
        \rowcolor{violet!12}[3.5\tabcolsep] \multirow{-2}{*}{Fanout} & \multirow{-2}{*}{Piroli et al.~\cite{piroli2024approximating}}& 1D line & \multirow{-2}{*}{$2n$} & \multirow{-2}{*}{$2$}  & \multirow{-2}{*}{$\frac{3}{2}n-2$} & $12$ & $7n-8$\\ \hline
        \rowcolor{violet!20}[3.5\tabcolsep] Fanout &  Fig.~\ref{fig:fanout_dyn}(b)(i) & 1D line & $2n+1$ & $2$ & $2n-1$ & $4$ & $4n-2$ \\ \hline
        \rowcolor{violet!20}[3.5\tabcolsep] Fanout & Fig.~\ref{fig:fanout_dyn}(b)(ii) & 1D line & $2n+1$ & $1$ & $n$ & $5$ & $3n-1$ \\ \hline
        \rowcolor{teal!10}[3.5\tabcolsep] CNOT ladder & Unitary & 1D line & $n+1$ & $0$ & $0$ & $n$ & $n$\\ \hline
        \rowcolor{teal!20}[3.5\tabcolsep] CNOT ladder & Fig.~\ref{fig:ladders}& 1D line & $2n+1$ & $1$ & $n$ & $2$ & $2n$\\ \hline
        \rowcolor{Green!10}[3.5\tabcolsep] Long-range CNOT & Unitary & 1D line & $n+1$ & $0$ & $0$ & $2n+(-1)^n$ & $4n-3$ \\ \hline
        \rowcolor{Green!20}[3.5\tabcolsep] Long-range CNOT & Fig.~\ref{fig:fanout_skip}(b)(ii) & 1D line & $2n+1$ & $1$ & $n$ & $7$ & $4n-2$ \\
    \end{tabular}
    \caption{Resources required for an implementation of the fan-out gate, as well as the CNOT ladder and the long-range CNOT gate on $n+1$ qubits. We have compiled all implementations to a 1D line for better comparison.}
    \label{tab:comparison}
    \end{ruledtabular}
\end{table*}

To emphasize the advantage of our constructions, let us review and compare our results to their solely unitary implementations, as well as different existing dynamic circuit protocols. Since the pulse duration of CNOT gates and mid-circuit measurements is an order of magnitude longer than that of single-qubit rotations \cite{iqp}, we only consider the CNOT depth and rounds of mid-circuit measurements within this resource analysis and throughout the rest of the paper. As the common ground in the protocols we are aware of is the quantum fan-out gate, we focus our comparison on the different implementations of this sub-routine, but we have also listed the resources required by our other protocols in Table~\ref{tab:comparison}.

Considering the unitary implementation of a quantum fan-out gate on $n+1$ qubits, we would not require any ancilla qubits. However, even with a star connectivity, meaning a one-to-all connectivity, the CNOT depth would be $n$. When transpiling the fan-out gate from star connectivity to a sparser connectivity, such as a 1D line, the depth and number of CNOT gates doubles.
The first dynamic circuit implementation to compare to is by Buhrman et al.~\cite{buhrman2023state}, which is using the Bell pair teleportation-based protocol for executing any Clifford circuits in constant depth as described in~\cite{jozsa2005introduction}. While it can be implemented on a 1D line, it requires $2n$ ancilla qubits and roughly $6n$ CNOT gates. The dynamic circuit implementation of Piroli et al.~\cite{piroli2024approximating} on the other hand requires only $n$ ancilla qubits and only roughly $2.5 n$ CNOT gates, but assumes a ladder topology for the qubits. With a 1D line a big overhead in terms of SWAP gates would be required, increasing both, the depth and the number of CNOT gates drastically. In our two proposed implementations, only $n$ ancilla qubits are required and they can be implemented on a 1D line. The first one, the fusion of two CNOT ladders, has a depth of $4$ CNOT layers and order of $4n$ CNOT gates, but requires two rounds of mid-circuit measurements. In contrast, the second proposal requires a depth of $5$ (not-dense) CNOT layers, roughly $3n$ CNOT gates in total, and only one round of mid-circuit measurements. Thus, we expect this last implementation to outperform all existing protocols and yield the best results.

\section{Experimental Results}
\label{sec:experiments}
To demonstrate the feasibility of our protocols, we have experimentally implemented the fan-out gate (Fig.~\ref{fig:fanout_dyn}(ii)) and the long-range CNOT gate (Fig.~\ref{fig:fanout_dyn}(b)~(ii)), comparing the unitary implementation with our measurement-based shallow implementation. All experiments were performed on the \texttt{ibm\_kyiv} device, which is one of the IBM Quantum Eagle processors, using Qiskit~\cite{qiskit2024}. The implementation details can be found in Appendix~\ref{app:expdetails}. Note that in these experiments we emulated dynamic circuits by adding a delay of $654$~ns corresponding to the classical processing time in which we did not apply any other pulses and applied the corrections in post-processing. 

\begin{figure}[htb]
\includegraphics[width=1.0\columnwidth]{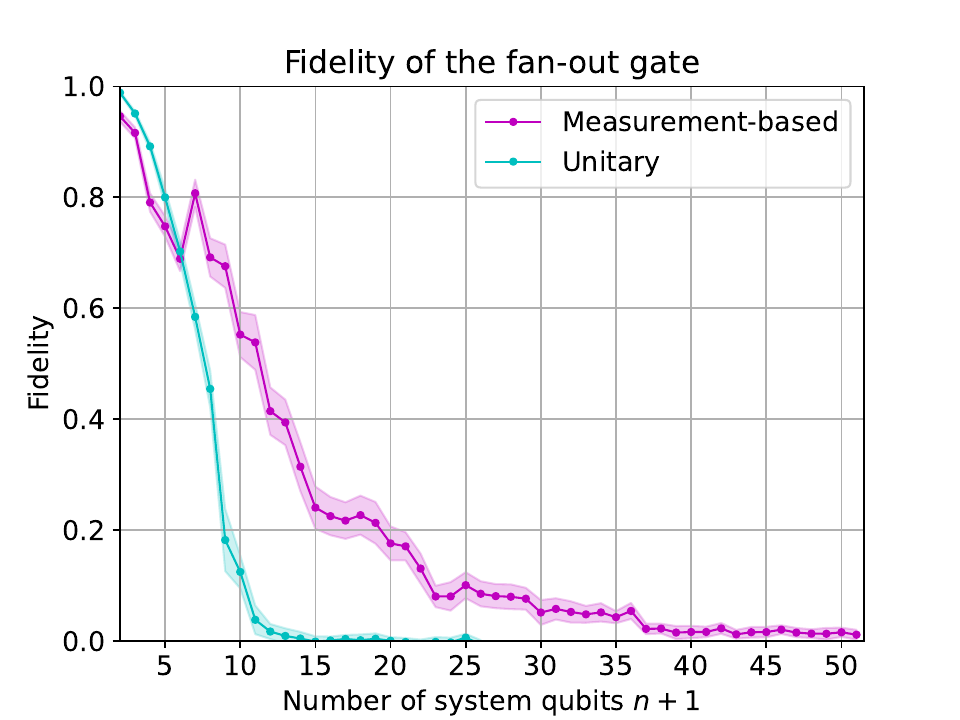}
\caption{Experimental results on \texttt{ibm\_kyiv} \cite{iqp} implementing the fan-out gate on $n+1$ qubits ($2n+1$ total qubits in the measurement-based implementation).
} 
\label{fig:results_fanout}
\end{figure}

\begin{figure}[htb]
\includegraphics[width=1.0\columnwidth]{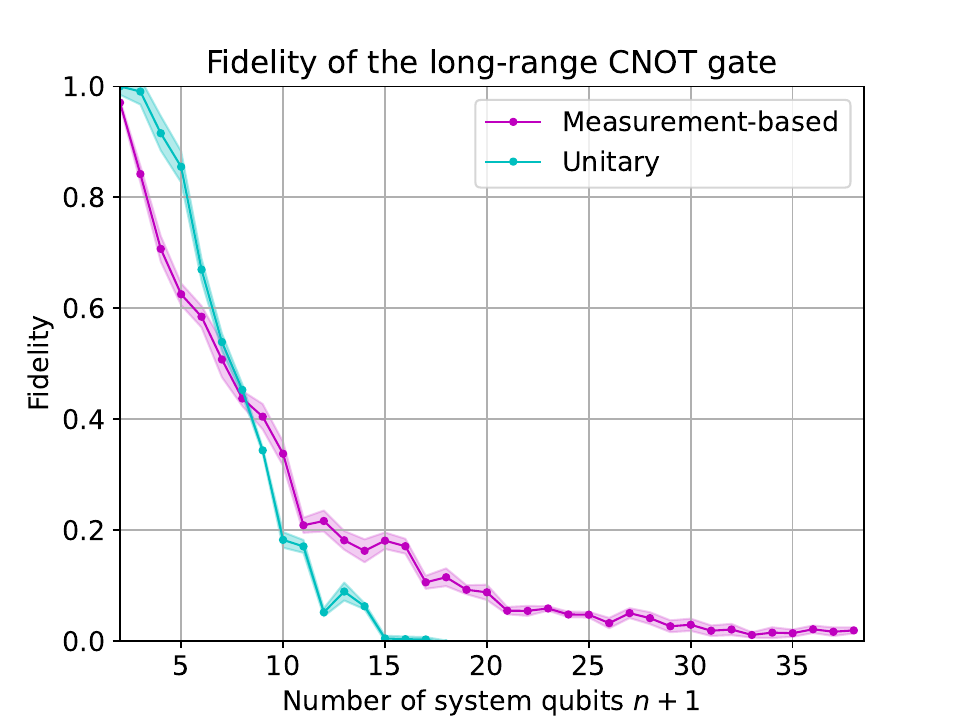}
\caption{Experimental results on \texttt{ibm\_kyiv} \cite{iqp} implementing the long-range CNOT gate on $n+1$ qubits ($2n+1$ total qubits in the measurement-based implementation).
} 
\label{fig:results_cnot}
\end{figure}

We estimated the fidelity of the full process for both operations. For the quantum fan-out gate, we consider the gate fidelity on the control qubit and all target qubits. For the long-range CNOT gate we consider the gate fidelity of the combination of a CNOT gate on the two outmost qubits and the identity on all other system qubits in-between. For more details on the fidelity estimation, see Appendix~\ref{app:fidelity}. The results are shown in Figs.~\ref{fig:results_fanout} and \ref{fig:results_cnot}, respectively, and include dynamical decoupling (DD)~\cite{Viola1999DD,Jurcevic2021} and measurement error mitigation on the final measurement outcomes of the system qubits only~\cite{Nation2021meas}. As expected, the measurement-based fidelities outperform the unitary implementations in both cases for larger $n$. While in the case of the fan-out gate (Fig.~\ref{fig:results_fanout}), the measurement-based protocol is beneficial for $n+1\geq 7$ (system) qubits, in the case of the long-range CNOT gate (Fig.~\ref{fig:results_cnot}) this holds for $n+1\geq 9$ (system) qubits. Due to the smaller depth and gate count, the overall fidelity is slightly higher for the fan-out gate, which allowed us to scale the measurement-based approach with non-zero fidelities to 51 system qubits, corresponding to a total of 101 qubits.

\section{Applications}
\label{sec:applications}

There is a wide range of applications benefiting from our constant depth implementations of the long-range entangling gates presented in Sec.~\ref{sec:ourimplementation}. 

Many applications of the quantum fan-out gate have already been described in~\cite{hoyer2005}, such as sorting, arithmetic operations, phase estimation, and the quantum Fourier transform, although not all of them might be feasible near-term. 
Additional applications of the quantum fan-out gate may include bit-flip symmetry verification~\cite{Shaydulin2021} and coherent Pauli checks~\cite{vandenBerg2023}.
Similarly, CNOT ladders are a common structure and appear, e.g., in the exponentiation of the excitation operators when implementing the UCCSD expansion \cite{Barkoutsos2018}. 
Further, long range gates can help to implement (trotterized) Hamiltonian simulation for applications with various--non-hardware-native--topologies. As we expect the latter to be among the most promising applications to be implemented on near-term hardware, we will discuss this in more depth in the following. 

\begin{figure}[htb]
\includegraphics[width=\columnwidth]{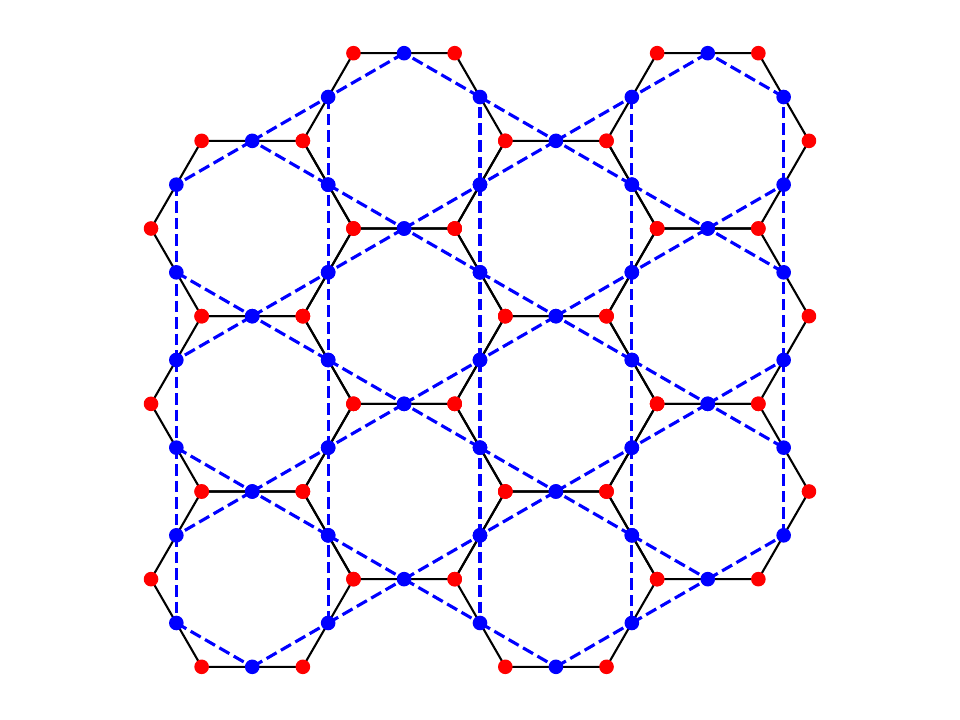}
\caption{
Choosing red nodes as state qubits and blue nodes as ancilla qubits allows to embed a hexagonal lattice into a heavy-hex topology. In contrast, choosing blue nodes as state qubits and red nodes as ancilla qubits leads to a Kagome lattice. The alternation of state and ancilla qubits allows to apply long-range gates between any pair of state qubits. As long as the connecting paths do not cross, multiple long range gates can also be applied in parallel.
}
\label{fig:topologies}
\end{figure}

In the following, we assume a quantum device with qubits arranged in a heavy-hex topology, cf.~eg.~\cite{iqp}, although other topologies allow similar constructions.
In order to apply the long-range gates introduced in this paper, we need alternating state and ancilla qubits. 
The heavy-hex lattice offers two natural choices that allow to embed state qubits on a hexagonal lattice as well as on a Kagome lattice, as illustrated in Fig.~\ref{fig:topologies}.

This allows us to achieve all-to-all connectivity between state qubits without any SWAP gates.
Furthermore, as long as the chosen paths to connect multiple pairs of qubits to implement long-range gates are not intersecting, those gates can even be applied in parallel.
In the context of Hamiltonian simulation, this allows us to implement a Trotter step for systems with heavy-hex or Kagome topologies with periodic boundary conditions in constant depth, by connecting opposite sides of the lattice via long-range gates.
The price to pay for the constant-depth long-range gates is that connecting neighboring state qubits now requires us to jump over the (clean) ancilla qubits, i.e., a single CNOT gate usually needs to be replaced by three CNOT gates. However, for larger systems, this is easily justified by the costs that would be introduced by implementing periodic boundary conditions via SWAP gates.

\begin{figure}[htb]
\includegraphics[width=0.6\columnwidth]{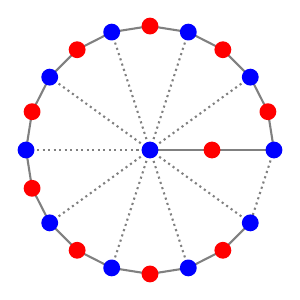}
\caption{Star and cart wheel topologies: A center state qubit surrounded by a ring of $n$ (here $n=10$) outer state qubits (all state qubits are blue) alternating with $n$ ancilla qubits. Connecting the center state qubit with all outer state qubits leads to the star topology. Additionally connecting all outer state qubits in a ring achieves the cart wheel topology.
Given a line of $2n+1$ qubits (indicated by the solid lines) we can implement, e.g., $R_{ZZ}$ gates on the star or cart wheel topology in constant depth, where the dotted lines indicate the connections that would be implemented by the new protocols introduced here.
}
\label{fig:cart_wheel_topology}
\end{figure}

An alternative topology that follows from our results is the star or cart wheel topology.
Suppose a ring of $n$ outer qubits around one center qubit, as illustrated in Fig.~\ref{fig:cart_wheel_topology}.
We can easily simulate Hamiltonians, i.e., implement Trotter steps, with such a topology in constant-depth on a line of $2n+1$ qubits leveraging the long-range gates introduced in the previous sections.
For instance, suppose we want to implement $R_{ZZ}(\theta)$ gates according to the graph in Fig.~\ref{fig:cart_wheel_topology}. 
We can achieve this in constant depth by first using the construction introduced in Sec.~\ref{sec:arbitraryU} to implement all gates between the central state qubit and all other state qubits in a single step, then applying all natively available gates (skipping the ancilla qubits), and last, applying one long-range gate to close the ring.
An exemplary application is the exciton transfer between an LH1 antenna complex and photosynthetic reaction center dimer~\cite{Pudlak2022}. 

\section{Conclusion and Outlook}
\label{sec:conclusion}
Using measurements and feed-forward operations can greatly reduce the circuit depth of various quantum sub-routines. Here, we show feasible constructions to implement CNOT-ladders, the quantum fan-out gate and the long-range CNOT gate in constant depth on a 1D line using one round of mid-circuit measurements and feed-forward operations, as well as some extensions like multi-qubit $R_{ZZ..Z}$ rotations and a fan-out gate with arbitrary parallel single-qubit unitaries that are controlled by a common qubit. We compare the resources of our fan-out protocol with previous ones, showing the increased feasibility, especially for hardware with sparse connectivity. To demonstrate the feasibility and the advantage of the measurement-based protocols, we also experimentally compare them with their unitary counterpart, where we see an increased performance for the measurement-based protocol when using more than 6 system qubits for the fan-out gate and when using more than 8 system qubits for the long-range CNOT gate. We also present some applications that could benefit using these protocols.\\

While we found different sub-routines that can be implemented in constant depth using mid-circuit measurements and feed-forward operations, it remains an open question to generalize such protocols and to find upper or lower bounds for the circuit depth as well as the number of required ancilla qubits of arbitrary Clifford or non-Clifford operations. 
Additionally, in future research, we might consider classically conditioned multi-qubit operations instead of only single-qubit operations. 

\acknowledgements
We thank Maika Takita, Almudena Carrera Vazquez, Daniel Egger, Jake Lishman, Diego Rist\`e and Jessie Yu for valuable discussions and support. We acknowledge the use of IBM Quantum services for this work. The views expressed are those of the authors, and do not reflect the official policy or position of IBM or the IBM Quantum team.

\newpage

\appendix
\onecolumngrid
\section{Derivations of Measurement-Based Gate Constructions}
\label{app:derivations}
\begin{figure}[htb]
\includegraphics[width=0.6\columnwidth]{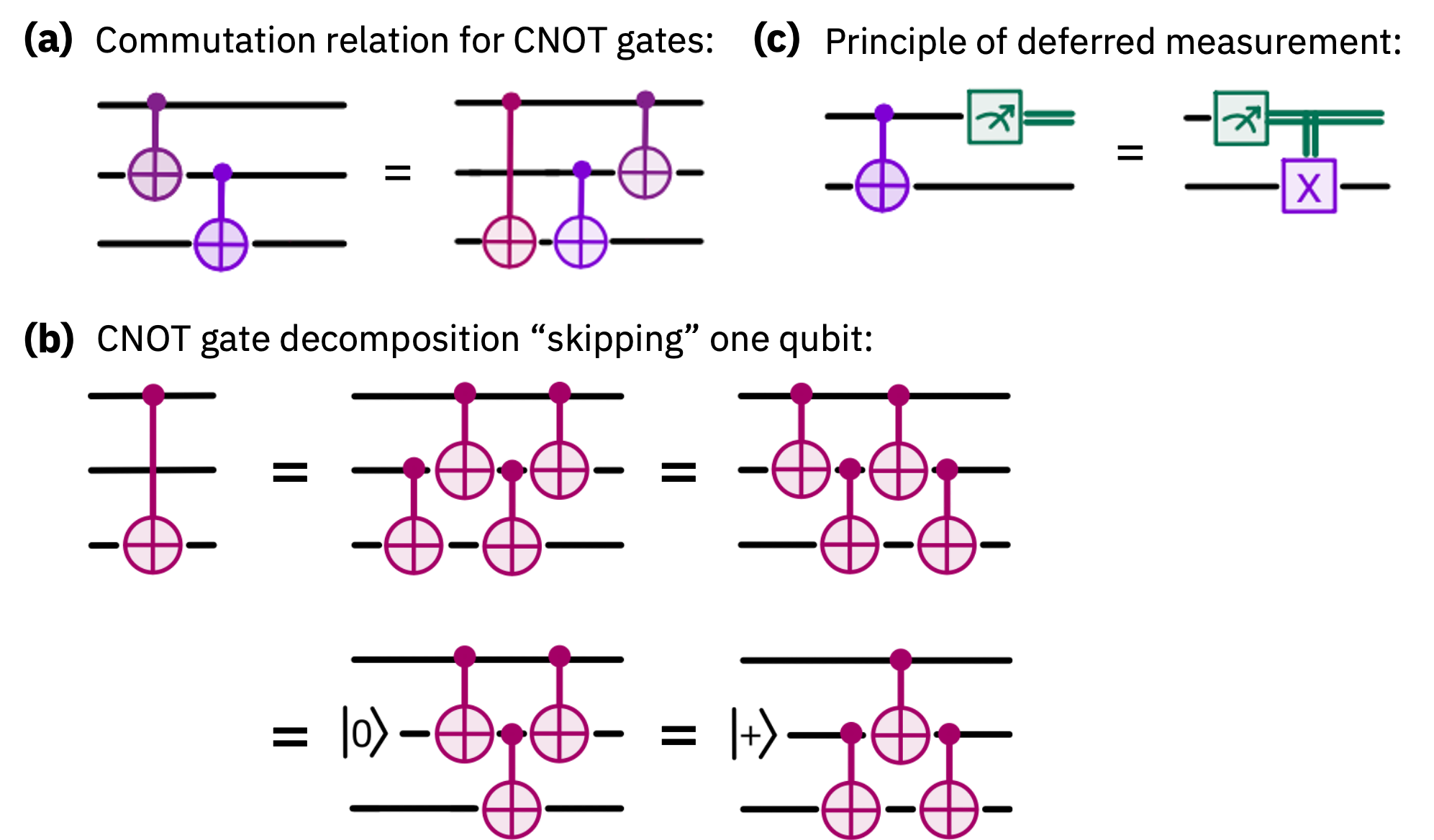}
\caption{Circuit equivalences that are used for deriving the measurement-based implementations.
} 
\label{fig:commrelations}
\end{figure}

In the following we derive some of the circuit equivalences from the main text. Let us start by describing some features that we will be using, that are illustrated in Fig.~\ref{fig:commrelations}:

\begin{enumerate}[(a)]
\item While CNOT gates commute when they are conditioned on the same qubit or have the same target qubit, we get an extra gate when they act on the same qubit differently and we change their order.
\item The commutation relation from (a) can be used to decompose a CNOT gate that ``skips" one qubit (assuming a 1D connectivity) into four CNOT gates on nearest neighbors. If the ``skipped" qubit is initially in state $\ket{0}$ or $\ket{+}$, one CNOT gate can be omitted.
\item \textit{Principle of deferred measurement}: a controlled gate followed by a measurement of the controlled qubit yields the same result as first performing the measurement and then applying a classically-controlled gate.
\end{enumerate}

\subsection{CNOT ladders}
\label{app:derivation_ladders}
\begin{figure}[htb]
\includegraphics[width=0.8\columnwidth]{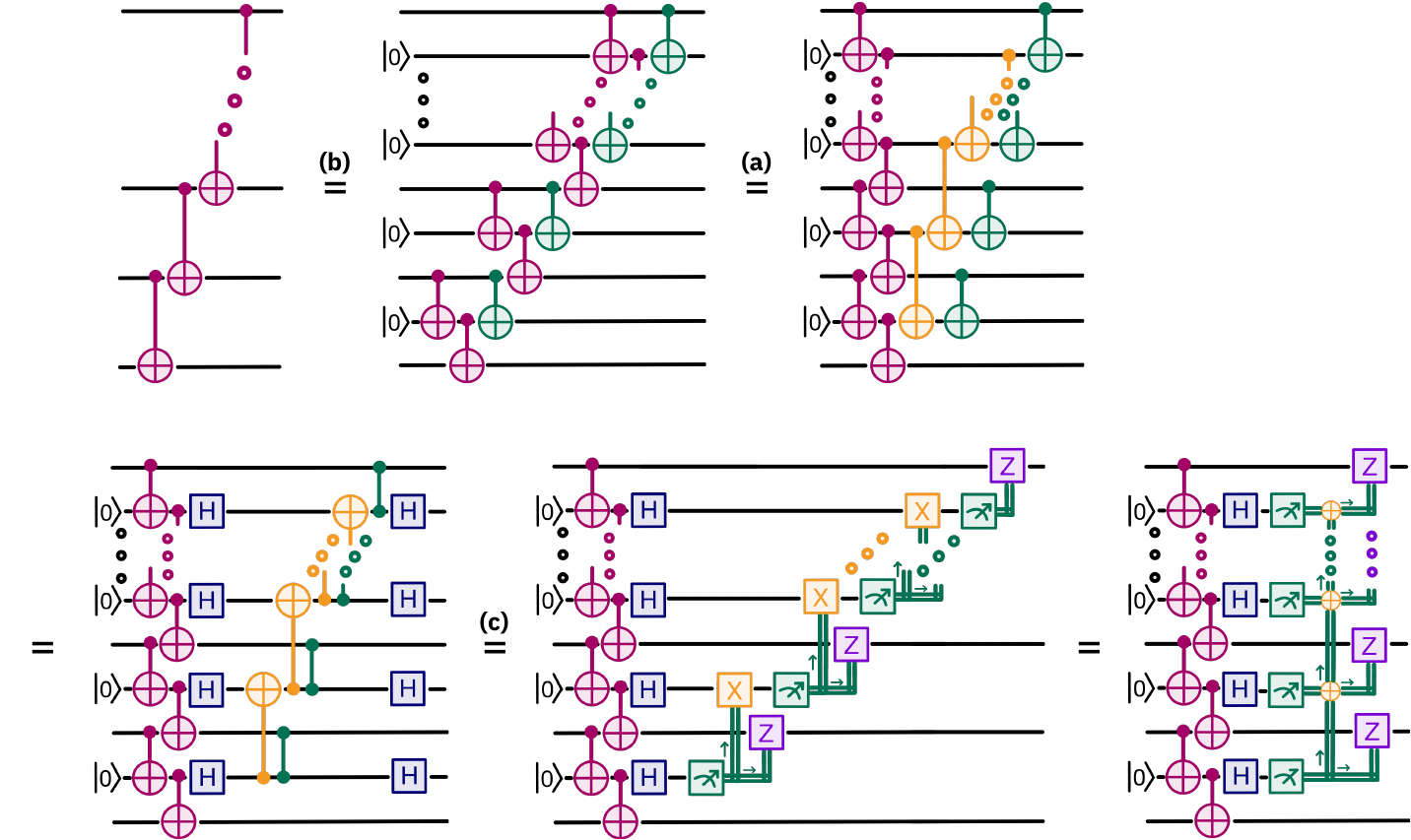}
\caption{Derivation of the measurement-based implementation of the CNOT ladders.
} 
\label{fig:ladders_derivation}
\end{figure}
The measurement-based implementation of the CNOT ladders is derived in Fig.~\ref{fig:ladders_derivation}. In a first step, empty ancilla qubits are inserted and ``skipped" by three CNOT gates (\ref{fig:commrelations}(b)). Then, some gates are moved which introduces the new orange gates according to \ref{fig:commrelations}(a). Inserting Hadamard gates before and after the orange and green gates changes the direction of those gates and allows to replace them by classically conditioned operations by applying the principle of deferred measurement (\ref{fig:commrelations}(c)). In a final step we simply use the fact that applying a bit flip just before measurement yields the same results as measuring first and then applying a classical bit flip.

\subsection{Fan-out gate}
\label{app:derivation_fanout}
\begin{figure}[htb]
\includegraphics[width=0.8\columnwidth]{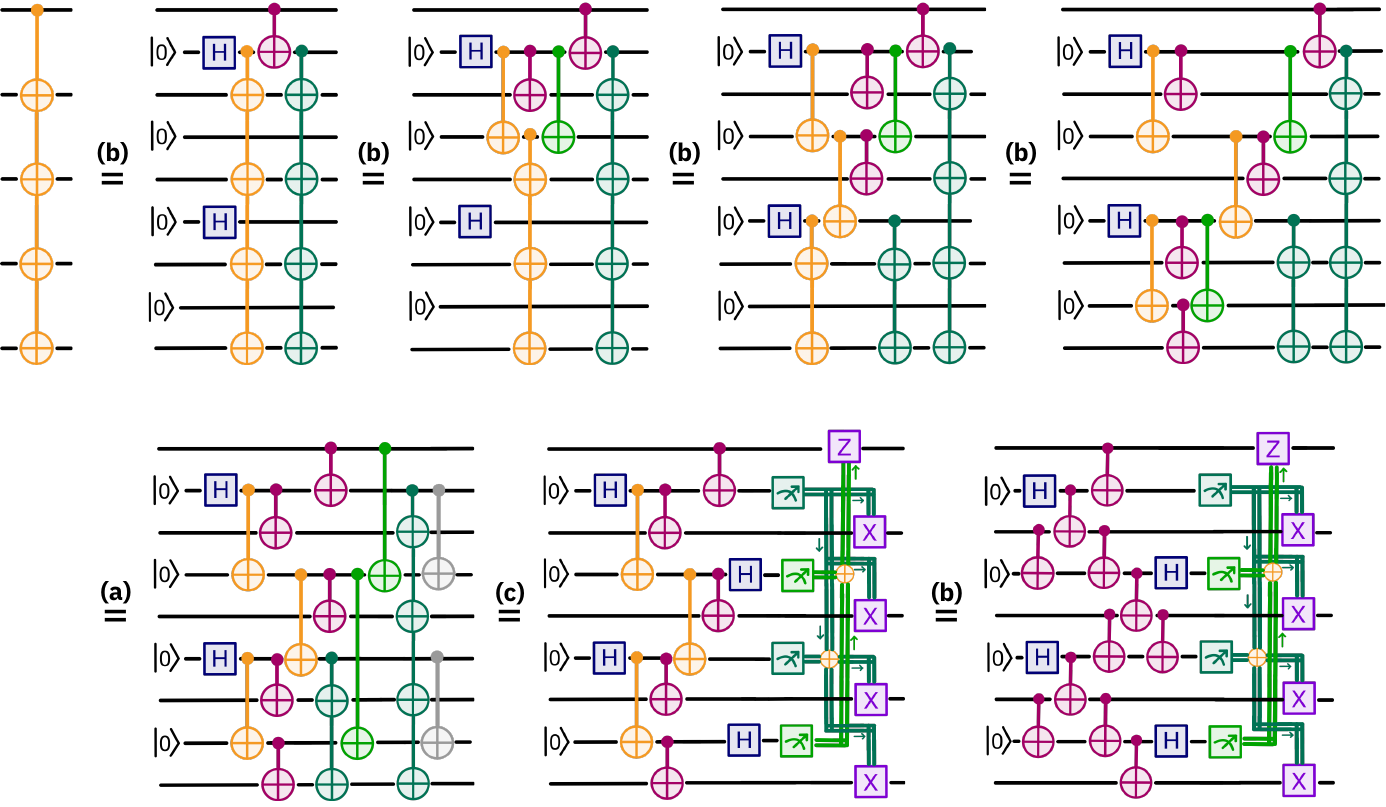}
\caption{Derivation of the measurement-based implementation of the quantum fan-out gate.
} 
\label{fig:fanout_derivation}
\end{figure}
The measurement-based implementation of the quantum fan-out gate is derived in Fig.~\ref{fig:fanout_derivation}. For simplicity, here we focus on $n=4$ target qubits, however, the derivation can be straightforwardly extended to arbitrary $n$. We start by inserting $n$ ancilla qubits with alternating initial states $\ket{+}$ and $\ket{0}$, respectively. In $n$ steps, each of them is used with the CNOT ``skipping" decomposition from (\ref{fig:commrelations}(b)) to decompose the long-range CNOT gates into smaller ones (first row of Fig.~\ref{fig:fanout_derivation}. Then the light green gates are pushed to the end, thereby resulting in additional light green gates. The original ones are colored grey as they do not need to be implemented as they only act between the final ancilla states, which are anyway measured and reset. Inserting Hadamard gates to change the direction of the light green gates allows to apply the principle of deferred measurement on all green gates and replace them by classically conditioned gates. In a final step, the orange CNOT gates need to be decomposed again into four nearest neighbor CNOT gates each according to~\ref{fig:commrelations}(b).

\subsection{Long-range CNOT gate}
\label{app:derivation_cnot}
\begin{figure}[htb]
\includegraphics[width=0.5\columnwidth]{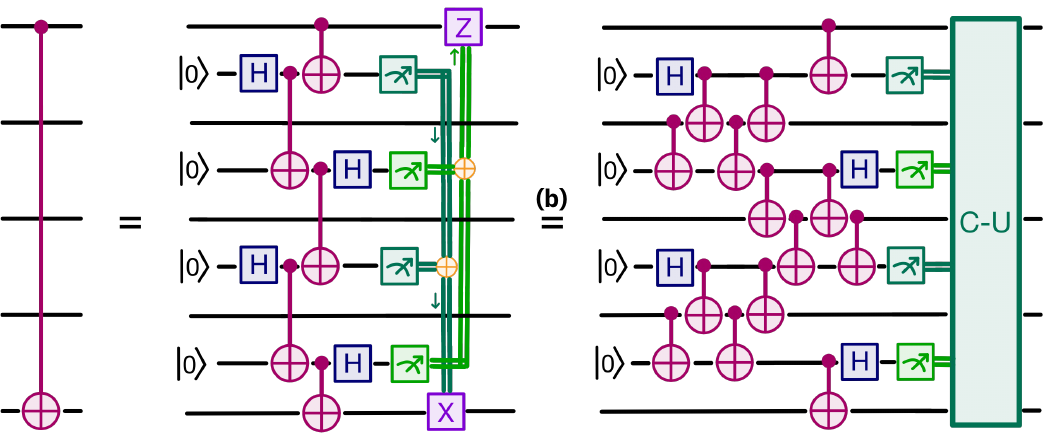}
\caption{Derivation of the measurement-based implementation of the long-range CNOT gate with system qubits along the way.
} 
\label{fig:cnot_derivation}
\end{figure}
The measurement-based implementation of the long-range CNOT gate is derived in Fig.~\ref{fig:cnot_derivation}. We start by inserting $n$ ancilla qubits and considering the measurement-based long-range CNOT gate teleportation protocol via those empty ancilla qubits, as e.g. explicitly derived in~\cite{baeumer2023}, thereby ``skipping'' the system qubits along the way. Next, we simply decompose those CNOT gates that ``skip" the system qubits into four nearest neighbor CNOT gates each, according to~\ref{fig:commrelations}(b). 

\subsection{State teleportation}
\label{app:derivation_teleportation}
\begin{figure}[htb]
\includegraphics[width=0.5\columnwidth]{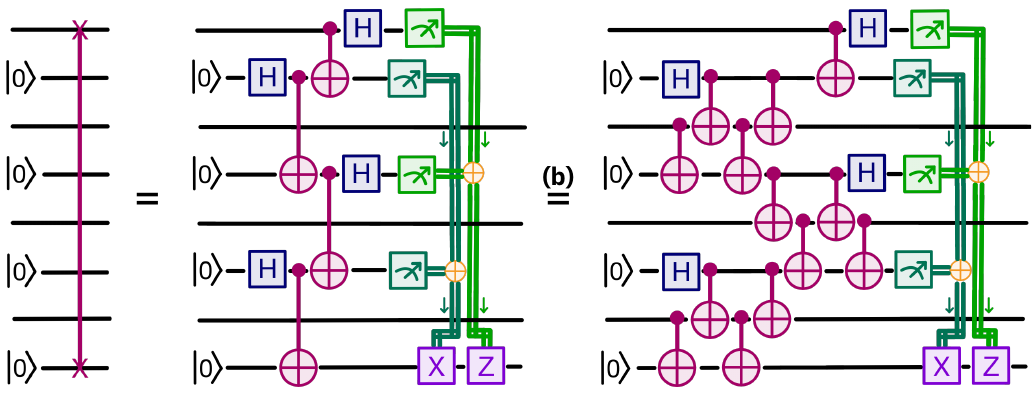}
\caption{Derivation of the measurement-based implementation of teleportation with system qubits along the way.
} 
\label{fig:teleportation}
\end{figure}
Similarly to the long-range CNOT implementation, the measurement-based teleportation can be implemented by inserting $n$ ancilla qubits and considering the standard teleportation protocol via those empty ancilla qubits, thereby ``skipping" the system qubits along the way, as shown in Fig.~\ref{fig:teleportation}. Next, we simply decompose those CNOT gates that ``skip" the system qubits into four nearest neighbor CNOT gates each, according to~\ref{fig:commrelations}(b). 

\section{Experimental details}
\label{app:expdetails}
\begin{figure}[htb]
\includegraphics[width=0.8\columnwidth]{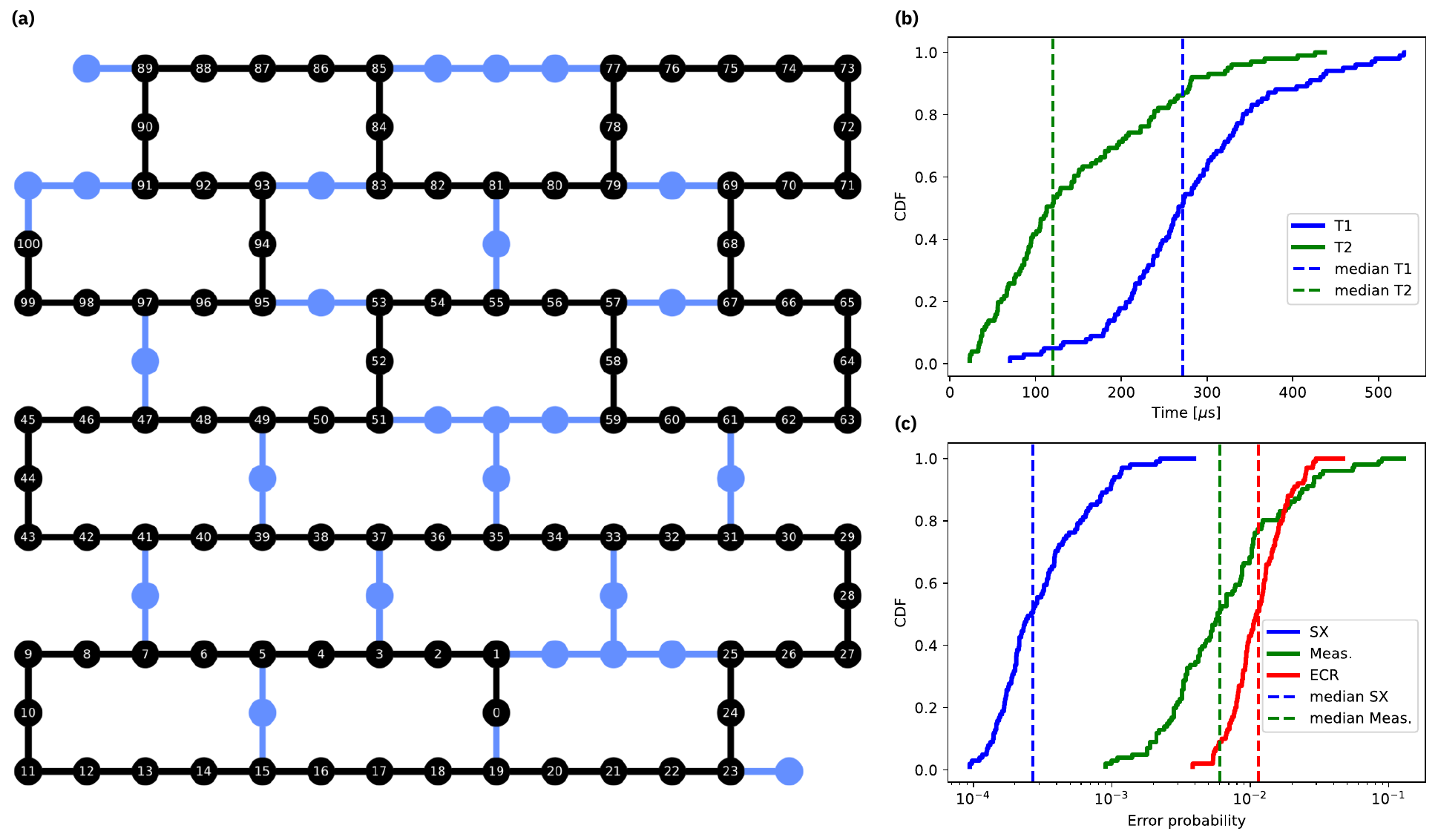}
\caption{Implementation details of the fan-out experiments. In (a), we show the device layout of \texttt{ibm\_kyiv}, with the $101$ qubits chosen for the measurement-based protocol marked in black. In (b) and (c), we plot the cumulative distribution of the T1 and T2 coherence times, the single qubit gate (SX), readout (Meas.) and two qubit echoed cross-resonance gate (ECR) error rates of the chosen qubits, as well as the corresponding median values.}
\label{fig:qubitdata_fanout}
\end{figure}

\begin{figure}[htb]
\includegraphics[width=0.8\columnwidth]{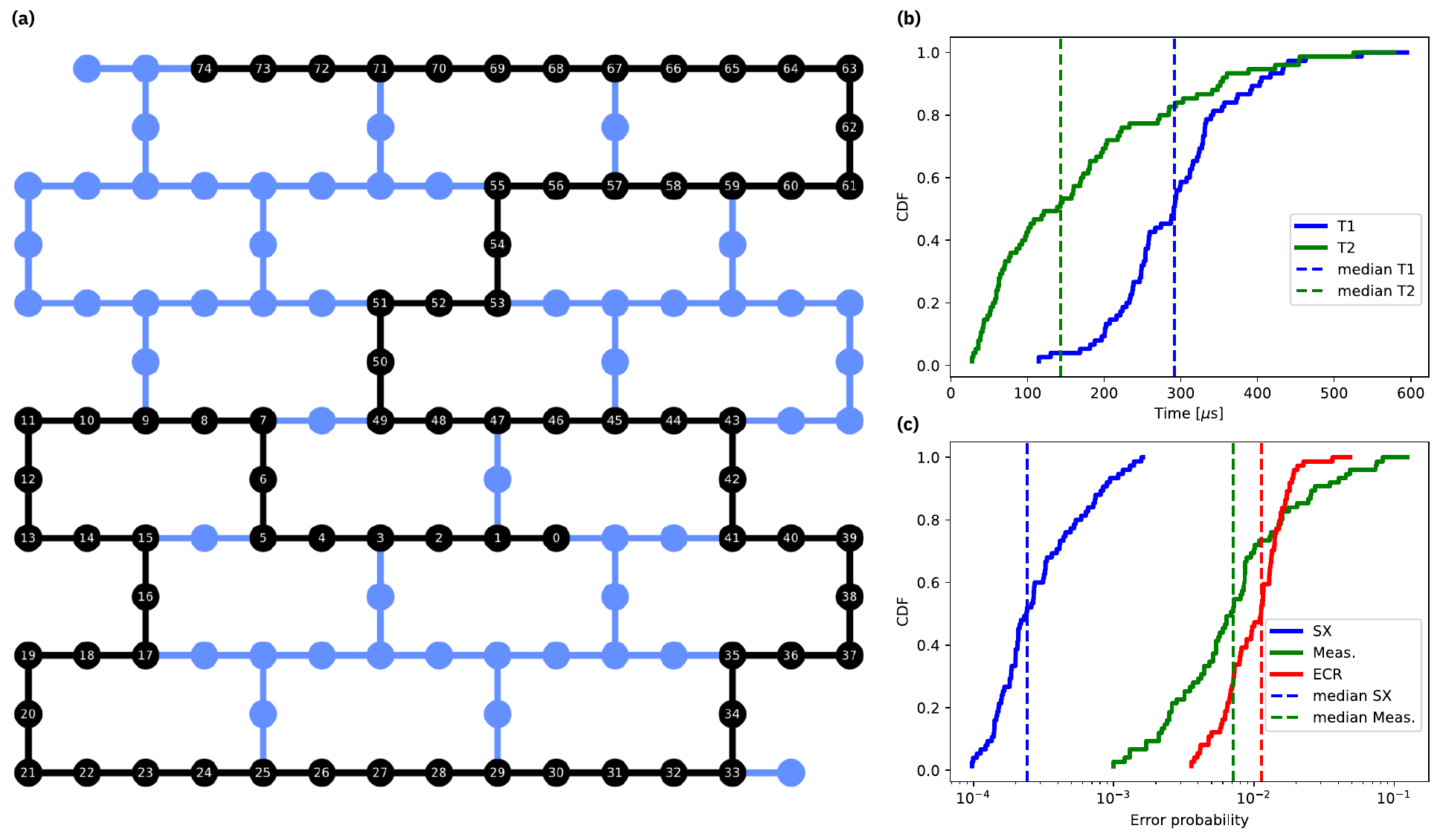}
\caption{Implementation details of the long-range cnot experiments. In (a), we show the device layout of \texttt{ibm\_kyiv}, with the $75$ qubits chosen for the measurement-based protocol marked in black. In (b) and (c), we plot the cumulative distribution of the T1 and T2 coherence times, the single qubit gate (SX), readout (Meas.) and two qubit echoed cross-resonance gate (ECR) error rates of the chosen qubits, as well as the corresponding median values.}
\label{fig:qubitdata_cnot}
\end{figure}

We perform all experiments on \texttt{ibm\_kyiv}, a 127-qubit superconducting IBM Quantum Eagle processor. The line of $101$ qubits chosen for the fan-out experiments are indicated in \cref{fig:qubitdata_fanout}(a) and line of $75$ qubits chosen for the long-range CNOT gate experiments in \cref{fig:qubitdata_cnot}(a). The cumulative distribution of their T1 and T2 coherence times, as well as of their different error rates are shown in \cref{fig:qubitdata_fanout}(b)-(c) and \cref{fig:qubitdata_cnot}(b)-(c), respectively, indicating also the corresponding median values. 
The two-qubit gate time is $0.56~\mu s$, the readout time $1.24~\mu s$ and the (simulated) classical processing time for feed-forward $0.65~\mu s$.

\section{Gate fidelity estimation}
\label{app:fidelity}

The gate fidelity between an ideal, noise-free gate~$\mathcal{U}(\rho_{\psi})\coloneqq U\rho_{\psi} U^{\dagger}$ and its experimental, noisy implementation~$\tilde{\mathcal{U}}(\rho_{\psi})\coloneqq \mathcal{U}(\Lambda(\rho_{\psi}))$, where $\Lambda$ is some effective noise channel and $\rho_{\psi}=\ket{\psi}\bra{\psi}$ is a quantum state, is defined as
the average state fidelity of the ideal and noisy output states, 
\begin{align}
\mathcal{F}_{\mathrm{gate}}\left(\mathcal{U},\tilde{\mathcal{U}}\right) &= \int\mathrm{d}\psi\,\mathcal{F}_{\mathrm{state}} \left(\mathcal{U}\left(\rho_{\psi}\right),\tilde{\mathcal{U}} \left(\rho_{\psi}\right)\right)\\
&= \int\mathrm{d}\psi\,\tr\left[\rho_{\psi}\Lambda \left(\rho_{\psi}\right)\right]\;,
\end{align}
where the integral is taken over the uniform Haar measure $\mathrm{d}\psi$ on state space~\cite{Nielsen2002} and $\mathcal{F}_{\mathrm{state}}(\rho_{\psi},\sigma)=\tr[\rho_{\psi}\sigma]$ is the Uhlmann-Jozsa state fidelity between the ideal pure quantum state $\rho_{\psi}$ and an arbitrary state $\sigma$. As the direct definition is not experimentally accessible, we use the following relation derived in \cite{Horodecki1999} to determine the gate fidelity via the process fidelity, which is more experimentally accessible:
\begin{align}
\mathcal{F}_\mathrm{gate}({{\mathcal{U}}}, {\tilde{\mathcal{U}}})
= 
\frac{d \cdot \mathcal{F}_\mathrm{proc}
({{\mathcal{U}}}, {\tilde{\mathcal{U}}})+1}{d+1}\;,
\end{align}
where $d=2^n$ is the dimension of the state space that the operator is acting on. It is connected via the Choi-Jamiolkowski isomophism~\cite{Jamiolkowski1972}, which maps any quantum operation $\Lambda$ on a $d$-dimensional space to its Choi state $\rho_\Lambda = (\mathbb{I}\otimes\Lambda)\ket{\phi}\bra{\phi}$, where $\ket{\phi} = \frac{1}{\sqrt{d}}\sum_{i=1}^d\ket{i}\otimes\ket{i}$. The process fidelity $\mathcal{F}_\mathrm{proc}
({{\mathcal{U}}}, {\tilde{\mathcal{U}}})$ is given by the state fidelity of the respective Choi states $\rho_{\mathcal{U}}$ and $\rho_{\tilde{\mathcal{U}}}$, which we can further decompose in terms of the Pauli decomposition as:
\begin{align}
\mathcal{F}_\mathrm{proc}
({{\mathcal{U}}}, {\tilde{\mathcal{U}}})
&:= \mathcal{F}_{\mathrm{state}}(\rho_{{\mathcal{U}}}, \rho_{\tilde{\mathcal{U}}}) 
= \tr \left[\rho_{{\mathcal{U}}} \rho_{\tilde{\mathcal{U}}} \right] 
= \sum_{i,j} \frac{\tr[\rho_{{\mathcal{U}}} (P_i\otimes P_j)] \tr[\rho_{\tilde{\mathcal{U}}} (P_i\otimes P_j)] }{d^2}
= \sum_{i,j} \frac{\langle P_{ij}\rangle^2_{\rho_{{\mathcal{U}}}}}{d^2}\frac{\langle P_{ij} \rangle_{\rho_{\tilde{\mathcal{U}}}}}{\langle P_{ij} \rangle_{\rho_{{\mathcal{U}}}}} \\
&= \sum_{i,j: \langle P_{ij} \rangle_{\rho_{{\mathcal{U}}}} \neq 0} r(P_{ij}) \frac{\langle P_{ij} \rangle_{\rho_{\tilde{\mathcal{U}}}}}{\langle P_{ij} \rangle_{\rho_{{\mathcal{U}}}}},
\end{align}
where $\langle P_{ij} \rangle_{\rho_{\tilde{\mathcal{U}}}}= \tr[\rho_{\tilde{\mathcal{U}}} (P_i \otimes P_j)]$ is an experimentally obtained expectation value of the Pauli operator $P_i \otimes P_j$ with $P_i, P_j \in\{\mathbb{I},\sigma_X,\sigma_Y,\sigma_Z\}^{\otimes n}$ and $\langle P_{ij} \rangle_{\rho_{{\mathcal{U}}}}= \tr[\rho_{{\mathcal{U}}} (P_i\otimes P_j)]$ the ideal one that can be theoretically calculated and from which we can determine the \textit{relevance distribution} $r(P_{ij}):=\frac{\langle P_{ij} \rangle^2_{\rho_{{\mathcal{U}}}}}{d^2}$.
Instead of a direct implementation of $\rho_{\tilde{\mathcal{U}}}=(\mathbb{I}\otimes\tilde{\mathcal{U}})\ket{\phi}\bra{\phi}$ followed by measuring random Pauli operators on all $2n$ qubits, we follow the more practical approach described in Ref.~\cite{daSilva2011}, where $\tilde{\mathcal{U}}$ is applied to the complex conjugate of a random product of eigenstates of local Pauli operators $P_i$, followed by a measurement of random Pauli operators $P_j$, which both act on $n$ qubits each, leading to the same expectation values 
\begin{align}
\langle P_{ij} \rangle_{\rho_{\tilde{\mathcal{U}}}}&:=\tr\left[(P_i \otimes P_j)\rho_{\tilde{\mathcal{U}}}\right]= \tr\left[P_j \tilde{\mathcal{U}} P_i^\ast\right]/d^2\;.
\end{align} 
For a Clifford operation on $n$ qubits, for each of the $4^n$ Pauli operators $P_i$ there is exactly one $P_j$ that yields a non-zero expectation value and for that $\langle P_{ij} \rangle_{\rho_{{\mathcal{U}}}} \in \{+1,-1\}$. Thus, the relevance distribution is uniform amongst those with $r(P_{ij})=\frac{1}{4^n}$.

As the number of non-zero expectation values scales exponentially, we employ the \textit{Monte Carlo state certification} method~\cite{daSilva2011,Flammia2011} to determine the fidelities.
It allows to sample $m$ random operators $\{P_{k}\}_{k=1..m}$ according to the relevance distribution $r(P_k)$ and determine their expectation values $\sigma_{k}$ to estimate the fidelity $\tilde{F} := \sum_{k=1}^m \frac{\langle P_{k} \rangle_{\rho_{\tilde{\mathcal{U}}}}}{\langle P_{k} \rangle_{\rho_{{\mathcal{U}}}}}$ which approximates the actual fidelity $F$ with an uncertainty that decreases as $\frac{1}{\sqrt{m}}$.
Note that there is also an uncertainty in estimating each $\sigma_{k}$, where for an additive precision~$\epsilon$ roughly $(\epsilon \rho_{k})^{-2}$ shots are required.


\begin{thebibliography}{28}%
\makeatletter
\providecommand \@ifxundefined [1]{%
 \@ifx{#1\undefined}
}%
\providecommand \@ifnum [1]{%
 \ifnum #1\expandafter \@firstoftwo
 \else \expandafter \@secondoftwo
 \fi
}%
\providecommand \@ifx [1]{%
 \ifx #1\expandafter \@firstoftwo
 \else \expandafter \@secondoftwo
 \fi
}%
\providecommand \natexlab [1]{#1}%
\providecommand \enquote  [1]{``#1''}%
\providecommand \bibnamefont  [1]{#1}%
\providecommand \bibfnamefont [1]{#1}%
\providecommand \citenamefont [1]{#1}%
\providecommand \href@noop [0]{\@secondoftwo}%
\providecommand \href [0]{\begingroup \@sanitize@url \@href}%
\providecommand \@href[1]{\@@startlink{#1}\@@href}%
\providecommand \@@href[1]{\endgroup#1\@@endlink}%
\providecommand \@sanitize@url [0]{\catcode `\\12\catcode `\$12\catcode
  `\&12\catcode `\#12\catcode `\^12\catcode `\_12\catcode `\%12\relax}%
\providecommand \@@startlink[1]{}%
\providecommand \@@endlink[0]{}%
\providecommand \url  [0]{\begingroup\@sanitize@url \@url }%
\providecommand \@url [1]{\endgroup\@href {#1}{\urlprefix }}%
\providecommand \urlprefix  [0]{URL }%
\providecommand \Eprint [0]{\href }%
\providecommand \doibase [0]{https://doi.org/}%
\providecommand \selectlanguage [0]{\@gobble}%
\providecommand \bibinfo  [0]{\@secondoftwo}%
\providecommand \bibfield  [0]{\@secondoftwo}%
\providecommand \translation [1]{[#1]}%
\providecommand \BibitemOpen [0]{}%
\providecommand \bibitemStop [0]{}%
\providecommand \bibitemNoStop [0]{.\EOS\space}%
\providecommand \EOS [0]{\spacefactor3000\relax}%
\providecommand \BibitemShut  [1]{\csname bibitem#1\endcsname}%
\let\auto@bib@innerbib\@empty
\bibitem [{\citenamefont {Kim}\ \emph {et~al.}(2023)\citenamefont {Kim},
  \citenamefont {Eddins}, \citenamefont {Anand}, \citenamefont {Wei},
  \citenamefont {van~den Berg}, \citenamefont {Rosenblatt}, \citenamefont
  {Nayfeh}, \citenamefont {Wu}, \citenamefont {Zaletel}, \citenamefont
  {Temme},\ and\ \citenamefont {Kandala}}]{kim2023utility}%
  \BibitemOpen
  \bibfield  {author} {\bibinfo {author} {\bibfnamefont {Y.}~\bibnamefont
  {Kim}}, \bibinfo {author} {\bibfnamefont {A.}~\bibnamefont {Eddins}},
  \bibinfo {author} {\bibfnamefont {S.}~\bibnamefont {Anand}}, \bibinfo
  {author} {\bibfnamefont {K.~X.}\ \bibnamefont {Wei}}, \bibinfo {author}
  {\bibfnamefont {E.}~\bibnamefont {van~den Berg}}, \bibinfo {author}
  {\bibfnamefont {S.}~\bibnamefont {Rosenblatt}}, \bibinfo {author}
  {\bibfnamefont {H.}~\bibnamefont {Nayfeh}}, \bibinfo {author} {\bibfnamefont
  {Y.}~\bibnamefont {Wu}}, \bibinfo {author} {\bibfnamefont {M.}~\bibnamefont
  {Zaletel}}, \bibinfo {author} {\bibfnamefont {K.}~\bibnamefont {Temme}},\
  and\ \bibinfo {author} {\bibfnamefont {A.}~\bibnamefont {Kandala}},\
  }\bibfield  {title} {\bibinfo {title} {Evidence for the utility of quantum
  computing before fault tolerance},\ }\href
  {https://doi.org/10.1038/s41586-023-06096-3} {\bibfield  {journal} {\bibinfo
  {journal} {Nature}\ }\textbf {\bibinfo {volume} {618}},\ \bibinfo {pages}
  {500} (\bibinfo {year} {2023})}\BibitemShut {NoStop}%
\bibitem [{\citenamefont {Malz}\ \emph {et~al.}(2024)\citenamefont {Malz},
  \citenamefont {Styliaris}, \citenamefont {Wei},\ and\ \citenamefont
  {Cirac}}]{Malz2024}%
  \BibitemOpen
  \bibfield  {author} {\bibinfo {author} {\bibfnamefont {D.}~\bibnamefont
  {Malz}}, \bibinfo {author} {\bibfnamefont {G.}~\bibnamefont {Styliaris}},
  \bibinfo {author} {\bibfnamefont {Z.-Y.}\ \bibnamefont {Wei}},\ and\ \bibinfo
  {author} {\bibfnamefont {J.~I.}\ \bibnamefont {Cirac}},\ }\bibfield  {title}
  {\bibinfo {title} {Preparation of matrix product states with log-depth
  quantum circuits},\ }\href {https://doi.org/10.1103/PhysRevLett.132.040404}
  {\bibfield  {journal} {\bibinfo  {journal} {Phys. Rev. Lett.}\ }\textbf
  {\bibinfo {volume} {132}},\ \bibinfo {pages} {040404} (\bibinfo {year}
  {2024})}\BibitemShut {NoStop}%
\bibitem [{\citenamefont {Sahay}\ and\ \citenamefont
  {Verresen}(2024)}]{sahay2024finitedepthpreparationtensornetwork}%
  \BibitemOpen
  \bibfield  {author} {\bibinfo {author} {\bibfnamefont {R.}~\bibnamefont
  {Sahay}}\ and\ \bibinfo {author} {\bibfnamefont {R.}~\bibnamefont
  {Verresen}},\ }\href {https://arxiv.org/abs/2404.17087} {\bibinfo {title}
  {Finite-depth preparation of tensor network states from measurement}}
  (\bibinfo {year} {2024}),\ \Eprint {https://arxiv.org/abs/2404.17087}
  {arXiv:2404.17087 [quant-ph]} \BibitemShut {NoStop}%
\bibitem [{\citenamefont {Stephen}\ and\ \citenamefont
  {Hart}(2024)}]{stephen2024preparingmatrixproductstates}%
  \BibitemOpen
  \bibfield  {author} {\bibinfo {author} {\bibfnamefont {D.~T.}\ \bibnamefont
  {Stephen}}\ and\ \bibinfo {author} {\bibfnamefont {O.}~\bibnamefont {Hart}},\
  }\href {https://arxiv.org/abs/2404.16360} {\bibinfo {title} {Preparing matrix
  product states via fusion: constraints and extensions}} (\bibinfo {year}
  {2024}),\ \Eprint {https://arxiv.org/abs/2404.16360} {arXiv:2404.16360
  [quant-ph]} \BibitemShut {NoStop}%
\bibitem [{\citenamefont {Zhang}\ \emph {et~al.}(2024)\citenamefont {Zhang},
  \citenamefont {Gopalakrishnan},\ and\ \citenamefont
  {Styliaris}}]{zhang2024characterizingmpspepspreparable}%
  \BibitemOpen
  \bibfield  {author} {\bibinfo {author} {\bibfnamefont {Y.}~\bibnamefont
  {Zhang}}, \bibinfo {author} {\bibfnamefont {S.}~\bibnamefont
  {Gopalakrishnan}},\ and\ \bibinfo {author} {\bibfnamefont {G.}~\bibnamefont
  {Styliaris}},\ }\href {https://arxiv.org/abs/2405.09615} {\bibinfo {title}
  {Characterizing mps and peps preparable via measurement and feedback}}
  (\bibinfo {year} {2024}),\ \Eprint {https://arxiv.org/abs/2405.09615}
  {arXiv:2405.09615 [quant-ph]} \BibitemShut {NoStop}%
\bibitem [{\citenamefont {Iqbal}\ \emph {et~al.}(2023)\citenamefont {Iqbal},
  \citenamefont {Tantivasadakarn}, \citenamefont {Gatterman}, \citenamefont
  {Gerber}, \citenamefont {Gilmore}, \citenamefont {Gresh}, \citenamefont
  {Hankin}, \citenamefont {Hewitt}, \citenamefont {Horst}, \citenamefont
  {Matheny}, \citenamefont {Mengle}, \citenamefont {Neyenhuis}, \citenamefont
  {Vishwanath}, \citenamefont {Foss-Feig}, \citenamefont {Verresen},\ and\
  \citenamefont {Dreyer}}]{iqbal2023}%
  \BibitemOpen
  \bibfield  {author} {\bibinfo {author} {\bibfnamefont {M.}~\bibnamefont
  {Iqbal}}, \bibinfo {author} {\bibfnamefont {N.}~\bibnamefont
  {Tantivasadakarn}}, \bibinfo {author} {\bibfnamefont {T.~M.}\ \bibnamefont
  {Gatterman}}, \bibinfo {author} {\bibfnamefont {J.~A.}\ \bibnamefont
  {Gerber}}, \bibinfo {author} {\bibfnamefont {K.}~\bibnamefont {Gilmore}},
  \bibinfo {author} {\bibfnamefont {D.}~\bibnamefont {Gresh}}, \bibinfo
  {author} {\bibfnamefont {A.}~\bibnamefont {Hankin}}, \bibinfo {author}
  {\bibfnamefont {N.}~\bibnamefont {Hewitt}}, \bibinfo {author} {\bibfnamefont
  {C.~V.}\ \bibnamefont {Horst}}, \bibinfo {author} {\bibfnamefont
  {M.}~\bibnamefont {Matheny}}, \bibinfo {author} {\bibfnamefont
  {T.}~\bibnamefont {Mengle}}, \bibinfo {author} {\bibfnamefont
  {B.}~\bibnamefont {Neyenhuis}}, \bibinfo {author} {\bibfnamefont
  {A.}~\bibnamefont {Vishwanath}}, \bibinfo {author} {\bibfnamefont
  {M.}~\bibnamefont {Foss-Feig}}, \bibinfo {author} {\bibfnamefont
  {R.}~\bibnamefont {Verresen}},\ and\ \bibinfo {author} {\bibfnamefont
  {H.}~\bibnamefont {Dreyer}},\ }\href {https://arxiv.org/abs/2302.01917}
  {\bibinfo {title} {Topological order from measurements and feed-forward on a
  trapped ion quantum computer}} (\bibinfo {year} {2023}),\ \Eprint
  {https://arxiv.org/abs/2302.01917} {arXiv:2302.01917 [quant-ph]} \BibitemShut
  {NoStop}%
\bibitem [{\citenamefont {Foss-Feig}\ \emph {et~al.}(2023)\citenamefont
  {Foss-Feig}, \citenamefont {Tikku}, \citenamefont {Lu}, \citenamefont
  {Mayer}, \citenamefont {Iqbal}, \citenamefont {Gatterman}, \citenamefont
  {Gerber}, \citenamefont {Gilmore}, \citenamefont {Gresh}, \citenamefont
  {Hankin}, \citenamefont {Hewitt}, \citenamefont {Horst}, \citenamefont
  {Matheny}, \citenamefont {Mengle}, \citenamefont {Neyenhuis}, \citenamefont
  {Dreyer}, \citenamefont {Hayes}, \citenamefont {Hsieh},\ and\ \citenamefont
  {Kim}}]{fossfeig2023}%
  \BibitemOpen
  \bibfield  {author} {\bibinfo {author} {\bibfnamefont {M.}~\bibnamefont
  {Foss-Feig}}, \bibinfo {author} {\bibfnamefont {A.}~\bibnamefont {Tikku}},
  \bibinfo {author} {\bibfnamefont {T.-C.}\ \bibnamefont {Lu}}, \bibinfo
  {author} {\bibfnamefont {K.}~\bibnamefont {Mayer}}, \bibinfo {author}
  {\bibfnamefont {M.}~\bibnamefont {Iqbal}}, \bibinfo {author} {\bibfnamefont
  {T.~M.}\ \bibnamefont {Gatterman}}, \bibinfo {author} {\bibfnamefont {J.~A.}\
  \bibnamefont {Gerber}}, \bibinfo {author} {\bibfnamefont {K.}~\bibnamefont
  {Gilmore}}, \bibinfo {author} {\bibfnamefont {D.}~\bibnamefont {Gresh}},
  \bibinfo {author} {\bibfnamefont {A.}~\bibnamefont {Hankin}}, \bibinfo
  {author} {\bibfnamefont {N.}~\bibnamefont {Hewitt}}, \bibinfo {author}
  {\bibfnamefont {C.~V.}\ \bibnamefont {Horst}}, \bibinfo {author}
  {\bibfnamefont {M.}~\bibnamefont {Matheny}}, \bibinfo {author} {\bibfnamefont
  {T.}~\bibnamefont {Mengle}}, \bibinfo {author} {\bibfnamefont
  {B.}~\bibnamefont {Neyenhuis}}, \bibinfo {author} {\bibfnamefont
  {H.}~\bibnamefont {Dreyer}}, \bibinfo {author} {\bibfnamefont
  {D.}~\bibnamefont {Hayes}}, \bibinfo {author} {\bibfnamefont {T.~H.}\
  \bibnamefont {Hsieh}},\ and\ \bibinfo {author} {\bibfnamefont {I.~H.}\
  \bibnamefont {Kim}},\ }\href {https://arxiv.org/abs/2302.03029} {\bibinfo
  {title} {Experimental demonstration of the advantage of adaptive quantum
  circuits}} (\bibinfo {year} {2023}),\ \Eprint
  {https://arxiv.org/abs/2302.03029} {arXiv:2302.03029 [quant-ph]} \BibitemShut
  {NoStop}%
\bibitem [{\citenamefont {Bäumer}\ \emph {et~al.}(2023)\citenamefont
  {Bäumer}, \citenamefont {Tripathi}, \citenamefont {Wang}, \citenamefont
  {Rall}, \citenamefont {Chen}, \citenamefont {Majumder}, \citenamefont
  {Seif},\ and\ \citenamefont {Minev}}]{baeumer2023}%
  \BibitemOpen
  \bibfield  {author} {\bibinfo {author} {\bibfnamefont {E.}~\bibnamefont
  {Bäumer}}, \bibinfo {author} {\bibfnamefont {V.}~\bibnamefont {Tripathi}},
  \bibinfo {author} {\bibfnamefont {D.~S.}\ \bibnamefont {Wang}}, \bibinfo
  {author} {\bibfnamefont {P.}~\bibnamefont {Rall}}, \bibinfo {author}
  {\bibfnamefont {E.~H.}\ \bibnamefont {Chen}}, \bibinfo {author}
  {\bibfnamefont {S.}~\bibnamefont {Majumder}}, \bibinfo {author}
  {\bibfnamefont {A.}~\bibnamefont {Seif}},\ and\ \bibinfo {author}
  {\bibfnamefont {Z.~K.}\ \bibnamefont {Minev}},\ }\href
  {https://arxiv.org/abs/2308.13065} {\bibinfo {title} {Efficient long-range
  entanglement using dynamic circuits}} (\bibinfo {year} {2023}),\ \Eprint
  {https://arxiv.org/abs/2308.13065} {arXiv:2308.13065 [quant-ph]} \BibitemShut
  {NoStop}%
\bibitem [{\citenamefont {Chen}\ \emph {et~al.}(2023)\citenamefont {Chen},
  \citenamefont {Zhu}, \citenamefont {Verresen}, \citenamefont {Seif},
  \citenamefont {Bäumer}, \citenamefont {Layden}, \citenamefont
  {Tantivasadakarn}, \citenamefont {Zhu}, \citenamefont {Sheldon},
  \citenamefont {Vishwanath}, \citenamefont {Trebst},\ and\ \citenamefont
  {Kandala}}]{chen2023nishimori}%
  \BibitemOpen
  \bibfield  {author} {\bibinfo {author} {\bibfnamefont {E.~H.}\ \bibnamefont
  {Chen}}, \bibinfo {author} {\bibfnamefont {G.-Y.}\ \bibnamefont {Zhu}},
  \bibinfo {author} {\bibfnamefont {R.}~\bibnamefont {Verresen}}, \bibinfo
  {author} {\bibfnamefont {A.}~\bibnamefont {Seif}}, \bibinfo {author}
  {\bibfnamefont {E.}~\bibnamefont {Bäumer}}, \bibinfo {author} {\bibfnamefont
  {D.}~\bibnamefont {Layden}}, \bibinfo {author} {\bibfnamefont
  {N.}~\bibnamefont {Tantivasadakarn}}, \bibinfo {author} {\bibfnamefont
  {G.}~\bibnamefont {Zhu}}, \bibinfo {author} {\bibfnamefont {S.}~\bibnamefont
  {Sheldon}}, \bibinfo {author} {\bibfnamefont {A.}~\bibnamefont {Vishwanath}},
  \bibinfo {author} {\bibfnamefont {S.}~\bibnamefont {Trebst}},\ and\ \bibinfo
  {author} {\bibfnamefont {A.}~\bibnamefont {Kandala}},\ }\href
  {https://arxiv.org/abs/2309.02863} {\bibinfo {title} {Realizing the nishimori
  transition across the error threshold for constant-depth quantum circuits}}
  (\bibinfo {year} {2023}),\ \Eprint {https://arxiv.org/abs/2309.02863}
  {arXiv:2309.02863 [quant-ph]} \BibitemShut {NoStop}%
\bibitem [{\citenamefont {Jozsa}(2005)}]{jozsa2005introduction}%
  \BibitemOpen
  \bibfield  {author} {\bibinfo {author} {\bibfnamefont {R.}~\bibnamefont
  {Jozsa}},\ }\href@noop {} {\bibinfo {title} {An introduction to measurement
  based quantum computation}} (\bibinfo {year} {2005}),\ \Eprint
  {https://arxiv.org/abs/quant-ph/0508124} {arXiv:quant-ph/0508124 [quant-ph]}
  \BibitemShut {NoStop}%
\bibitem [{\citenamefont {Buhrman}\ \emph {et~al.}(2023)\citenamefont
  {Buhrman}, \citenamefont {Folkertsma}, \citenamefont {Loff},\ and\
  \citenamefont {Neumann}}]{buhrman2023state}%
  \BibitemOpen
  \bibfield  {author} {\bibinfo {author} {\bibfnamefont {H.}~\bibnamefont
  {Buhrman}}, \bibinfo {author} {\bibfnamefont {M.}~\bibnamefont {Folkertsma}},
  \bibinfo {author} {\bibfnamefont {B.}~\bibnamefont {Loff}},\ and\ \bibinfo
  {author} {\bibfnamefont {N.~M.~P.}\ \bibnamefont {Neumann}},\ }\href@noop {}
  {\bibinfo {title} {State preparation by shallow circuits using feed forward}}
  (\bibinfo {year} {2023}),\ \Eprint {https://arxiv.org/abs/2307.14840}
  {arXiv:2307.14840 [quant-ph]} \BibitemShut {NoStop}%
\bibitem [{\citenamefont {Piroli}\ \emph {et~al.}(2024)\citenamefont {Piroli},
  \citenamefont {Styliaris},\ and\ \citenamefont
  {Cirac}}]{piroli2024approximating}%
  \BibitemOpen
  \bibfield  {author} {\bibinfo {author} {\bibfnamefont {L.}~\bibnamefont
  {Piroli}}, \bibinfo {author} {\bibfnamefont {G.}~\bibnamefont {Styliaris}},\
  and\ \bibinfo {author} {\bibfnamefont {J.~I.}\ \bibnamefont {Cirac}},\
  }\href@noop {} {\bibinfo {title} {Approximating many-body quantum states with
  quantum circuits and measurements}} (\bibinfo {year} {2024}),\ \Eprint
  {https://arxiv.org/abs/2403.07604} {arXiv:2403.07604 [quant-ph]} \BibitemShut
  {NoStop}%
\bibitem [{\citenamefont {Quantum}(2024)}]{iqp}%
  \BibitemOpen
  \bibfield  {author} {\bibinfo {author} {\bibfnamefont {I.}~\bibnamefont
  {Quantum}},\ }\href@noop {} {\bibinfo {title} {{IBM Quantum Platform}}},\
  \bibinfo {howpublished} {\url{https://quantum.ibm.com}} (\bibinfo {year}
  {2024}),\ \bibinfo {note} {[Online; accessed 2-July-2024]}\BibitemShut
  {NoStop}%
\bibitem [{\citenamefont {H{\o}yer}\ and\ \citenamefont {{\v
  S}palek}(2005)}]{hoyer2005}%
  \BibitemOpen
  \bibfield  {author} {\bibinfo {author} {\bibfnamefont {P.}~\bibnamefont
  {H{\o}yer}}\ and\ \bibinfo {author} {\bibfnamefont {R.}~\bibnamefont {{\v
  S}palek}},\ }\bibfield  {title} {\bibinfo {title} {Quantum fan-out is
  powerful},\ }\href {https://doi.org/10.4086/toc.2005.v001a005} {\bibfield
  {journal} {\bibinfo  {journal} {Theory of Computing}\ }\textbf {\bibinfo
  {volume} {1}},\ \bibinfo {pages} {81} (\bibinfo {year} {2005})}\BibitemShut
  {NoStop}%
\bibitem [{\citenamefont {Nielsen}\ and\ \citenamefont
  {Chuang}(2011)}]{NielsenChuang}%
  \BibitemOpen
  \bibfield  {author} {\bibinfo {author} {\bibfnamefont {M.~A.}\ \bibnamefont
  {Nielsen}}\ and\ \bibinfo {author} {\bibfnamefont {I.~L.}\ \bibnamefont
  {Chuang}},\ }\href
  {https://www.amazon.com/Quantum-Computation-Information-10th-Anniversary/dp/1107002176?SubscriptionId=AKIAIOBINVZYXZQZ2U3A&tag=chimbori05-20&linkCode=xm2&camp=2025&creative=165953&creativeASIN=1107002176}
  {\emph {\bibinfo {title} {Quantum Computation and Quantum Information: 10th
  Anniversary Edition}}}\ (\bibinfo  {publisher} {Cambridge University Press},\
  \bibinfo {year} {2011})\BibitemShut {NoStop}%
\bibitem [{\citenamefont {Javadi-Abhari}\ \emph {et~al.}(2024)\citenamefont
  {Javadi-Abhari}, \citenamefont {Treinish}, \citenamefont {Krsulich},
  \citenamefont {Wood}, \citenamefont {Lishman}, \citenamefont {Gacon},
  \citenamefont {Martiel}, \citenamefont {Nation}, \citenamefont {Bishop},
  \citenamefont {Cross}, \citenamefont {Johnson},\ and\ \citenamefont
  {Gambetta}}]{qiskit2024}%
  \BibitemOpen
  \bibfield  {author} {\bibinfo {author} {\bibfnamefont {A.}~\bibnamefont
  {Javadi-Abhari}}, \bibinfo {author} {\bibfnamefont {M.}~\bibnamefont
  {Treinish}}, \bibinfo {author} {\bibfnamefont {K.}~\bibnamefont {Krsulich}},
  \bibinfo {author} {\bibfnamefont {C.~J.}\ \bibnamefont {Wood}}, \bibinfo
  {author} {\bibfnamefont {J.}~\bibnamefont {Lishman}}, \bibinfo {author}
  {\bibfnamefont {J.}~\bibnamefont {Gacon}}, \bibinfo {author} {\bibfnamefont
  {S.}~\bibnamefont {Martiel}}, \bibinfo {author} {\bibfnamefont {P.~D.}\
  \bibnamefont {Nation}}, \bibinfo {author} {\bibfnamefont {L.~S.}\
  \bibnamefont {Bishop}}, \bibinfo {author} {\bibfnamefont {A.~W.}\
  \bibnamefont {Cross}}, \bibinfo {author} {\bibfnamefont {B.~R.}\ \bibnamefont
  {Johnson}},\ and\ \bibinfo {author} {\bibfnamefont {J.~M.}\ \bibnamefont
  {Gambetta}},\ }\href {https://doi.org/10.48550/arXiv.2405.08810} {\bibinfo
  {title} {Quantum computing with {Q}iskit}} (\bibinfo {year} {2024}),\ \Eprint
  {https://arxiv.org/abs/2405.08810} {arXiv:2405.08810 [quant-ph]} \BibitemShut
  {NoStop}%
\bibitem [{\citenamefont {Viola}\ \emph {et~al.}(1999)\citenamefont {Viola},
  \citenamefont {Knill},\ and\ \citenamefont {Lloyd}}]{Viola1999DD}%
  \BibitemOpen
  \bibfield  {author} {\bibinfo {author} {\bibfnamefont {L.}~\bibnamefont
  {Viola}}, \bibinfo {author} {\bibfnamefont {E.}~\bibnamefont {Knill}},\ and\
  \bibinfo {author} {\bibfnamefont {S.}~\bibnamefont {Lloyd}},\ }\bibfield
  {title} {\bibinfo {title} {Dynamical decoupling of open quantum systems},\
  }\href {https://doi.org/10.1103/PhysRevLett.82.2417} {\bibfield  {journal}
  {\bibinfo  {journal} {Phys. Rev. Lett.}\ }\textbf {\bibinfo {volume} {82}},\
  \bibinfo {pages} {2417} (\bibinfo {year} {1999})}\BibitemShut {NoStop}%
\bibitem [{\citenamefont {Jurcevic}\ \emph {et~al.}(2021)\citenamefont
  {Jurcevic}, \citenamefont {Javadi-Abhari}, \citenamefont {Bishop},
  \citenamefont {Lauer}, \citenamefont {Bogorin}, \citenamefont {Brink},
  \citenamefont {Capelluto}, \citenamefont {G\"{u}nl\"{u}k}, \citenamefont
  {Itoko}, \citenamefont {Kanazawa}, \citenamefont {Kandala}, \citenamefont
  {Keefe}, \citenamefont {Krsulich}, \citenamefont {Landers}, \citenamefont
  {Lewandowski}, \citenamefont {McClure}, \citenamefont {Nannicini},
  \citenamefont {Narasgond}, \citenamefont {Nayfeh}, \citenamefont {Pritchett},
  \citenamefont {Rothwell}, \citenamefont {Srinivasan}, \citenamefont
  {Sundaresan}, \citenamefont {Wang}, \citenamefont {Wei}, \citenamefont
  {Wood}, \citenamefont {Yau}, \citenamefont {Zhang}, \citenamefont {Dial},
  \citenamefont {Chow},\ and\ \citenamefont {Gambetta}}]{Jurcevic2021}%
  \BibitemOpen
  \bibfield  {author} {\bibinfo {author} {\bibfnamefont {P.}~\bibnamefont
  {Jurcevic}}, \bibinfo {author} {\bibfnamefont {A.}~\bibnamefont
  {Javadi-Abhari}}, \bibinfo {author} {\bibfnamefont {L.~S.}\ \bibnamefont
  {Bishop}}, \bibinfo {author} {\bibfnamefont {I.}~\bibnamefont {Lauer}},
  \bibinfo {author} {\bibfnamefont {D.~F.}\ \bibnamefont {Bogorin}}, \bibinfo
  {author} {\bibfnamefont {M.}~\bibnamefont {Brink}}, \bibinfo {author}
  {\bibfnamefont {L.}~\bibnamefont {Capelluto}}, \bibinfo {author}
  {\bibfnamefont {O.}~\bibnamefont {G\"{u}nl\"{u}k}}, \bibinfo {author}
  {\bibfnamefont {T.}~\bibnamefont {Itoko}}, \bibinfo {author} {\bibfnamefont
  {N.}~\bibnamefont {Kanazawa}}, \bibinfo {author} {\bibfnamefont
  {A.}~\bibnamefont {Kandala}}, \bibinfo {author} {\bibfnamefont {G.~A.}\
  \bibnamefont {Keefe}}, \bibinfo {author} {\bibfnamefont {K.}~\bibnamefont
  {Krsulich}}, \bibinfo {author} {\bibfnamefont {W.}~\bibnamefont {Landers}},
  \bibinfo {author} {\bibfnamefont {E.~P.}\ \bibnamefont {Lewandowski}},
  \bibinfo {author} {\bibfnamefont {D.~T.}\ \bibnamefont {McClure}}, \bibinfo
  {author} {\bibfnamefont {G.}~\bibnamefont {Nannicini}}, \bibinfo {author}
  {\bibfnamefont {A.}~\bibnamefont {Narasgond}}, \bibinfo {author}
  {\bibfnamefont {H.~M.}\ \bibnamefont {Nayfeh}}, \bibinfo {author}
  {\bibfnamefont {E.}~\bibnamefont {Pritchett}}, \bibinfo {author}
  {\bibfnamefont {M.~B.}\ \bibnamefont {Rothwell}}, \bibinfo {author}
  {\bibfnamefont {S.}~\bibnamefont {Srinivasan}}, \bibinfo {author}
  {\bibfnamefont {N.}~\bibnamefont {Sundaresan}}, \bibinfo {author}
  {\bibfnamefont {C.}~\bibnamefont {Wang}}, \bibinfo {author} {\bibfnamefont
  {K.~X.}\ \bibnamefont {Wei}}, \bibinfo {author} {\bibfnamefont {C.~J.}\
  \bibnamefont {Wood}}, \bibinfo {author} {\bibfnamefont {J.-B.}\ \bibnamefont
  {Yau}}, \bibinfo {author} {\bibfnamefont {E.~J.}\ \bibnamefont {Zhang}},
  \bibinfo {author} {\bibfnamefont {O.~E.}\ \bibnamefont {Dial}}, \bibinfo
  {author} {\bibfnamefont {J.~M.}\ \bibnamefont {Chow}},\ and\ \bibinfo
  {author} {\bibfnamefont {J.~M.}\ \bibnamefont {Gambetta}},\ }\bibfield
  {title} {\bibinfo {title} {Demonstration of quantum volume 64 on a
  superconducting quantum computing system},\ }\href
  {https://doi.org/10.1088/2058-9565/abe519} {\bibfield  {journal} {\bibinfo
  {journal} {Quantum Sci. Technol.}\ }\textbf {\bibinfo {volume} {6}},\
  \bibinfo {pages} {025020} (\bibinfo {year} {2021})}\BibitemShut {NoStop}%
\bibitem [{\citenamefont {Nation}\ \emph {et~al.}(2021)\citenamefont {Nation},
  \citenamefont {Kang}, \citenamefont {Sundaresan},\ and\ \citenamefont
  {Gambetta}}]{Nation2021meas}%
  \BibitemOpen
  \bibfield  {author} {\bibinfo {author} {\bibfnamefont {P.~D.}\ \bibnamefont
  {Nation}}, \bibinfo {author} {\bibfnamefont {H.}~\bibnamefont {Kang}},
  \bibinfo {author} {\bibfnamefont {N.}~\bibnamefont {Sundaresan}},\ and\
  \bibinfo {author} {\bibfnamefont {J.~M.}\ \bibnamefont {Gambetta}},\
  }\bibfield  {title} {\bibinfo {title} {Scalable mitigation of measurement
  errors on quantum computers},\ }\href
  {https://doi.org/10.1103/PRXQuantum.2.040326} {\bibfield  {journal} {\bibinfo
   {journal} {PRX Quantum}\ }\textbf {\bibinfo {volume} {2}},\ \bibinfo {pages}
  {040326} (\bibinfo {year} {2021})}\BibitemShut {NoStop}%
\bibitem [{\citenamefont {Shaydulin}\ and\ \citenamefont
  {Galda}(2021)}]{Shaydulin2021}%
  \BibitemOpen
  \bibfield  {author} {\bibinfo {author} {\bibfnamefont {R.}~\bibnamefont
  {Shaydulin}}\ and\ \bibinfo {author} {\bibfnamefont {A.}~\bibnamefont
  {Galda}},\ }\bibfield  {title} {\bibinfo {title} {Error mitigation for deep
  quantum optimization circuits by leveraging problem symmetries},\ }in\ \href
  {https://doi.org/10.1109/QCE52317.2021.00046} {\emph {\bibinfo {booktitle}
  {2021 IEEE International Conference on Quantum Computing and Engineering
  (QCE)}}}\ (\bibinfo {year} {2021})\ pp.\ \bibinfo {pages}
  {291--300}\BibitemShut {NoStop}%
\bibitem [{\citenamefont {van~den Berg}\ \emph {et~al.}(2023)\citenamefont
  {van~den Berg}, \citenamefont {Bravyi}, \citenamefont {Gambetta},
  \citenamefont {Jurcevic}, \citenamefont {Maslov},\ and\ \citenamefont
  {Temme}}]{vandenBerg2023}%
  \BibitemOpen
  \bibfield  {author} {\bibinfo {author} {\bibfnamefont {E.}~\bibnamefont
  {van~den Berg}}, \bibinfo {author} {\bibfnamefont {S.}~\bibnamefont
  {Bravyi}}, \bibinfo {author} {\bibfnamefont {J.~M.}\ \bibnamefont
  {Gambetta}}, \bibinfo {author} {\bibfnamefont {P.}~\bibnamefont {Jurcevic}},
  \bibinfo {author} {\bibfnamefont {D.}~\bibnamefont {Maslov}},\ and\ \bibinfo
  {author} {\bibfnamefont {K.}~\bibnamefont {Temme}},\ }\bibfield  {title}
  {\bibinfo {title} {Single-shot error mitigation by coherent pauli checks},\
  }\href {https://doi.org/10.1103/PhysRevResearch.5.033193} {\bibfield
  {journal} {\bibinfo  {journal} {Phys. Rev. Res.}\ }\textbf {\bibinfo {volume}
  {5}},\ \bibinfo {pages} {033193} (\bibinfo {year} {2023})}\BibitemShut
  {NoStop}%
\bibitem [{\citenamefont {Barkoutsos}\ \emph {et~al.}(2018)\citenamefont
  {Barkoutsos}, \citenamefont {Gonthier}, \citenamefont {Sokolov},
  \citenamefont {Moll}, \citenamefont {Salis}, \citenamefont {Fuhrer},
  \citenamefont {Ganzhorn}, \citenamefont {Egger}, \citenamefont {Troyer},
  \citenamefont {Mezzacapo}, \citenamefont {Filipp},\ and\ \citenamefont
  {Tavernelli}}]{Barkoutsos2018}%
  \BibitemOpen
  \bibfield  {author} {\bibinfo {author} {\bibfnamefont {P.~K.}\ \bibnamefont
  {Barkoutsos}}, \bibinfo {author} {\bibfnamefont {J.~F.}\ \bibnamefont
  {Gonthier}}, \bibinfo {author} {\bibfnamefont {I.}~\bibnamefont {Sokolov}},
  \bibinfo {author} {\bibfnamefont {N.}~\bibnamefont {Moll}}, \bibinfo {author}
  {\bibfnamefont {G.}~\bibnamefont {Salis}}, \bibinfo {author} {\bibfnamefont
  {A.}~\bibnamefont {Fuhrer}}, \bibinfo {author} {\bibfnamefont
  {M.}~\bibnamefont {Ganzhorn}}, \bibinfo {author} {\bibfnamefont {D.~J.}\
  \bibnamefont {Egger}}, \bibinfo {author} {\bibfnamefont {M.}~\bibnamefont
  {Troyer}}, \bibinfo {author} {\bibfnamefont {A.}~\bibnamefont {Mezzacapo}},
  \bibinfo {author} {\bibfnamefont {S.}~\bibnamefont {Filipp}},\ and\ \bibinfo
  {author} {\bibfnamefont {I.}~\bibnamefont {Tavernelli}},\ }\bibfield  {title}
  {\bibinfo {title} {Quantum algorithms for electronic structure calculations:
  Particle-hole hamiltonian and optimized wave-function expansions},\ }\href
  {https://doi.org/10.1103/PhysRevA.98.022322} {\bibfield  {journal} {\bibinfo
  {journal} {Phys. Rev. A}\ }\textbf {\bibinfo {volume} {98}},\ \bibinfo
  {pages} {022322} (\bibinfo {year} {2018})}\BibitemShut {NoStop}%
\bibitem [{\citenamefont {Pudlak}\ and\ \citenamefont
  {Pincak}(2022)}]{Pudlak2022}%
  \BibitemOpen
  \bibfield  {author} {\bibinfo {author} {\bibfnamefont {M.}~\bibnamefont
  {Pudlak}}\ and\ \bibinfo {author} {\bibfnamefont {R.}~\bibnamefont
  {Pincak}},\ }\href {https://doi.org/10.21203/rs.3.rs-2084477/v1} {\bibinfo
  {title} {Photosynthetic complex: exciton transfer and electron-hole
  separation}} (\bibinfo {year} {2022})\BibitemShut {NoStop}%
\bibitem [{\citenamefont {Nielsen}(2002)}]{Nielsen2002}%
  \BibitemOpen
  \bibfield  {author} {\bibinfo {author} {\bibfnamefont {M.~A.}\ \bibnamefont
  {Nielsen}},\ }\bibfield  {title} {\bibinfo {title} {A simple formula for the
  average gate fidelity of a quantum dynamical operation},\ }\href
  {https://doi.org/https://doi.org/10.1016/S0375-9601(02)01272-0} {\bibfield
  {journal} {\bibinfo  {journal} {Physics Letters A}\ }\textbf {\bibinfo
  {volume} {303}},\ \bibinfo {pages} {249} (\bibinfo {year}
  {2002})}\BibitemShut {NoStop}%
\bibitem [{\citenamefont {Horodecki}\ \emph {et~al.}(1999)\citenamefont
  {Horodecki}, \citenamefont {Horodecki},\ and\ \citenamefont
  {Horodecki}}]{Horodecki1999}%
  \BibitemOpen
  \bibfield  {author} {\bibinfo {author} {\bibfnamefont {M.}~\bibnamefont
  {Horodecki}}, \bibinfo {author} {\bibfnamefont {P.}~\bibnamefont
  {Horodecki}},\ and\ \bibinfo {author} {\bibfnamefont {R.}~\bibnamefont
  {Horodecki}},\ }\bibfield  {title} {\bibinfo {title} {General teleportation
  channel, singlet fraction, and quasidistillation},\ }\href
  {https://doi.org/10.1103/PhysRevA.60.1888} {\bibfield  {journal} {\bibinfo
  {journal} {Phys. Rev. A}\ }\textbf {\bibinfo {volume} {60}},\ \bibinfo
  {pages} {1888} (\bibinfo {year} {1999})}\BibitemShut {NoStop}%
\bibitem [{\citenamefont {Jamio{\l}kowski}(1972)}]{Jamiolkowski1972}%
  \BibitemOpen
  \bibfield  {author} {\bibinfo {author} {\bibfnamefont {A.}~\bibnamefont
  {Jamio{\l}kowski}},\ }\bibfield  {title} {\bibinfo {title} {Linear
  transformations which preserve trace and positive semidefiniteness of
  operators},\ }\href
  {https://doi.org/https://doi.org/10.1016/0034-4877(72)90011-0} {\bibfield
  {journal} {\bibinfo  {journal} {Rep. Math. Phys.}\ }\textbf {\bibinfo
  {volume} {3}},\ \bibinfo {pages} {275} (\bibinfo {year} {1972})}\BibitemShut
  {NoStop}%
\bibitem [{\citenamefont {da~Silva}\ \emph {et~al.}(2011)\citenamefont
  {da~Silva}, \citenamefont {Landon-Cardinal},\ and\ \citenamefont
  {Poulin}}]{daSilva2011}%
  \BibitemOpen
  \bibfield  {author} {\bibinfo {author} {\bibfnamefont {M.~P.}\ \bibnamefont
  {da~Silva}}, \bibinfo {author} {\bibfnamefont {O.}~\bibnamefont
  {Landon-Cardinal}},\ and\ \bibinfo {author} {\bibfnamefont {D.}~\bibnamefont
  {Poulin}},\ }\bibfield  {title} {\bibinfo {title} {Practical characterization
  of quantum devices without tomography},\ }\href
  {https://doi.org/10.1103/PhysRevLett.107.210404} {\bibfield  {journal}
  {\bibinfo  {journal} {Phys. Rev. Lett.}\ }\textbf {\bibinfo {volume} {107}},\
  \bibinfo {pages} {210404} (\bibinfo {year} {2011})}\BibitemShut {NoStop}%
\bibitem [{\citenamefont {Flammia}\ and\ \citenamefont
  {Liu}(2011)}]{Flammia2011}%
  \BibitemOpen
  \bibfield  {author} {\bibinfo {author} {\bibfnamefont {S.~T.}\ \bibnamefont
  {Flammia}}\ and\ \bibinfo {author} {\bibfnamefont {Y.-K.}\ \bibnamefont
  {Liu}},\ }\bibfield  {title} {\bibinfo {title} {Direct fidelity estimation
  from few pauli measurements},\ }\href
  {https://doi.org/10.1103/PhysRevLett.106.230501} {\bibfield  {journal}
  {\bibinfo  {journal} {Phys. Rev. Lett.}\ }\textbf {\bibinfo {volume} {106}},\
  \bibinfo {pages} {230501} (\bibinfo {year} {2011})}\BibitemShut {NoStop}%
\end{thebibliography}
\end{document}